\newif\ifspin
\spinfalse

\documentclass[aps,floatfix,twocolumn,showpacs,10pt,nofootinbib]{revtex4-1}
\usepackage[T1]{fontenc}
\usepackage{amsfonts}
\usepackage{hyperref}
\usepackage[usenames,dvipsnames]{color}
\newcommand{\be}{\begin{equation}}
\newcommand{\ee}{\end{equation}}
\setcitestyle{super}

\usepackage{amssymb,amsmath,graphicx}

\begin{document}


\title{Orbital effects of a strong in-plane magnetic field on a gate-defined quantum dot}


\author{Peter Stano$^{1,2,3}$, Chen-Hsuan Hsu$^{1}$, Leon C. Camenzind$^{4}$, Liuqi Yu$^{4}$, Dominik Zumb\"uhl$^{4}$, Daniel Loss$^{1,4}$}
\affiliation{$^{1}$RIKEN Center for Emergent Matter Science, 2-1 Hirosawa, Wako, Saitama, 351-0198 Japan}
\affiliation{$^{2}$Department of Applied Physics, School of Engineering, University of Tokyo, 7-3-1 Hongo, Bunkyo-ku, Tokyo 113-8656, Japan}
\affiliation{$^{3}$Institute of Physics, Slovak Academy of Sciences, 845 11 Bratislava, Slovakia}
\affiliation{$^{4}$Department of Physics, University of Basel, Klingelbergstrasse 82, CH-4056 Basel, Switzerland}

\date{\today}

\begin{abstract}
We theoretically investigate the orbital effects of an in-plane magnetic field on the spectrum of a quantum dot embedded in a two-dimensional electron gas (2DEG). We derive an effective two-dimensional Hamiltonian where these effects enter in proportion to the flux penetrating the 2DEG. We quantify the latter in detail for harmonic, triangular, and  square potential of the heterostructure.
We show how the orbital effects allow one to extract a wealth of information, for example, on the heterostructure interface, the quantum dot size and orientation, and the spin-orbit fields.
We illustrate the formalism by extracting this information from recent measured data [L.~C.~Camenzind, {\it et al.}, arXiv:1804.00162; Nat. Commun. {\bf 9}, 3454 (2018)].
\end{abstract}


\maketitle

\newcommand{\x}{\hat{\textbf{x}}}
\newcommand{\y}{\hat{\textbf{y}}}
\newcommand{\z}{\hat{\textbf{z}}}

\newcommand{\Bin}{\textbf{b}}
\newcommand{\Binscalar}{b}
\newcommand{\Binunit}{\hat{\textbf{b}}}
\newcommand{\Bz}{B_z}
\newcommand{\Bzvec}{B_z \z}
\newcommand{\Rin}{\textbf{r}}
\newcommand{\Rz}{z\z}
\newcommand{\Pin}{\textbf{p}}
\newcommand{\Ain}{\textbf{a}_{||}}
\newcommand{\Az}{\textbf{a}_{\perp}}

\newcommand{\ave}[2]{\overline{#1}^{#2}}

\newcommand{\flux}{\Phi}

 \section{Introduction}

The two-dimensional electron gas (2DEG) is a versatile platform for a host of devices and applications of nanotechnology.\cite{ando1982:RMP} In experiments with spin qubits realized in gated quantum dots based on 2DEGs,\cite{loss1999:PRA,hanson2007:RMP,kloeffel2013:ARCMP} it is usual to use magnetic fields which are applied parallel to the 2DEG plane (in-plane fields), and which are large, of the order of Tesla. The former is because, unless the quantum Hall effect physics is aimed at, strong orbital effects of the Lorentz force are undesirable. The second is because many tasks require an appreciable energy contrast (say, with respect to the temperature) of the spin opposite states, relying on the inherently small Zeeman splitting.

Necessarily, the assumption of the electron gas being quasi-two-dimensional breaks down once the in-plane field becomes too large, roughly when the magnetic length $\sqrt{\hbar /eB}$ becomes comparable to the width of the 2DEG. To give an example, for the 2DEG width of 8 nm, this occurs at about $10$ Tesla. The typical field strengths of a few Tesla are therefore not negligibly small compared to this crossover field, and one expects sizable effects which go beyond the quasi-two-dimensional model.\cite{zumbuhl2004:PRB,zumbuhl2005:PRB} Quantification of such orbital effects of an in-plane field on spectra of quantum dots is what we pursue here.

We find that these effects are well captured by a renormalization (increase) of the effective mass along the axis which lies within the 2DEG plane and is perpendicular to the magnetic field. We give the renormalization factor as a function of the flux corresponding to the in-plane component of the magnetic field penetrating an area expressed as a square of an effective 2DEG width. We relate the latter to the nominal width for 2DEGs with the most typical confinement profiles, namely harmonic, triangular, and rectangular.

We propose a two-dimensional effective model which remains reliable even for very large fields, well beyond the crossover field. The corresponding Hamiltonian is given in Eq.~\eqref{eq:main result} and it reduces the presence of the third dimension to a single parameter, the above mentioned effective 2DEG width. It gives essentially exact results if the magnetic field is purely in-plane and the heterostructure confinement is harmonic, and compares well with a fully 3D description in other cases, including an appreciable out-of-plane component of the magnetic field, which is, for example, typical for designs with micromagnets.\cite{otsuka2016:SR,yoneda2015:APE}

Perhaps the most important point we want to make in this work is that the orbital effects of in-plane fields should not be viewed as a nuisance, invalidating the simple model being a 2DEG with a zero width. Namely, as the direction of the external magnetic field can be experimentally well controlled, these effects can reveal the quantum dot orientation within the 2DEG plane, as well as its size in all three directions.\cite{yu} This, so far missing, spectroscopic tool is essential for a quantitative assessment of, for example, the spin-orbit fields,\cite{scarlino2014:PRL} or the hyperfine electron-nuclear interaction, and the related limits on the spin relaxation,\cite{camenzind,malkoc2016:PRB} dephasing,\cite{delbecq2016:PRL} or measurement fidelities.\cite{nakajima2017:PRL} To illustrate the power of these tools, we use them here to fit the strengths of the spin-orbit interactions in a GaAs quantum dot. We find excellent agreement with values extracted from an independent fit based on the directional variation of the spin relaxation time done in Ref.~\citenum{camenzind}. It demonstrates an unprecedented level of control over, and understanding of, spin qubits in quantum dots.

The paper is structured as follows. In Sec.~II, we introduce a three-dimensional effective-mass model of a quantum dot. In Sec.~III, we derive the effective 2D Hamiltonian which includes the effects of the in-plane field in the leading order by a perturbation theory. Here, we also give details on the effective width for various 2DEG profiles. In Sec.~IV, we discuss the effects expected in the dot spectra. In Sec.~V, we generalize the Hamiltonian beyond the perturbative regime of modest magnetic fields. In Sec.~VI we illustrate the usefulness of our results by extracting the 2DEG interfacial electric field from experimental data, with which one can calculate the spin-orbit fields.
Several auxiliary results are given in three appendices. Appendix A contains details on the matrix elements needed to convert the spectroscopic data to the heterostructure-interface characteristics. Appendix B gives, for reference, the spectrum of a general quadratic Hamiltonian, which then includes also our effective 2D Hamiltonian. Appendix C contains the evaluation of the formulas for the strengths of the linear spin-orbit interactions.

\section{Model}

We consider a quantum dot defined by gates on top of a two-dimensional electron gas created by a semiconductor heterostructure.  Since we are interested in effects which go beyond the lowest-order approximation, being that of a quasi-two-dimensional dot, we need a three-dimensional model to start with. The ${\bf k} \cdot {\bf p}$ theory based on the envelope-function approximation is an established method to obtain models which are simple enough for analytical calculations, yet reliable in treating the effects of the band structure and the sharp interface of the heterostructure.

\subsection{Zeroth order effective mass Hamiltonian}

\label{sec:notations}

The leading-order term for the conduction band of a zinc-blende semiconductor, such as GaAs, is
\be
H=\frac{ \textbf{P}^2 }{2m} + V({\bf R}).
\label{eq:H zeroth order}
\ee
It describes particles with a quadratic energy dispersion which move in the externally imposed confinement potential $V(\textbf{R})$, created by gates and the heterostructure composition. Here, $\textbf{R}$ is the three-dimensional position vector, and
\be
\textbf{P} = -i\hbar (\partial_x, \partial_y, \partial_z) + e\textbf{A},
\ee
is the canonical momentum, with $e$ the absolute value of the electron charge, and $\textbf{A}$ the vector potential of the magnetic field $\textbf{B}$, through which the orbital effects enter. On this level, the only effect of the crystal is that the effective parameter, the mass $m$, differs from the value of the electron mass in vacuum.

Before continuing, let us make a comment. Here, we analyze the magnetic field effects on the orbital energies of the dot. The magnetic field influences, similarly, the spin structure of the dot states. The latter effects are smaller than the former, analogously to the Zeeman energy being smaller than the orbital energy. We do not include the spin-dependent effects in Eq.~\eqref{eq:H zeroth order} and report on these elsewhere.\cite{stano2018:PRB}

\subsection{In-plane and perpendicular coordinates}

We assume that the heterostructure is grown along the [001] crystallographic axis, which is further called the perpendicular direction, with the unit vector $\z$, and the corresponding coordinate $z$. The remaining two crystallographic directions are denoted as $\x=[100]$ and $\y=[010]$, and we call them in plane. The separation to perpendicular and in-plane coordinates is motivated by strong anisotropy of the three-dimensional confinement. Namely, it is a sum of a harder perpendicular (heterostructure) part, $v(z)$, and a softer in-plane (quantum dot) part $V_{\rm 2D}(x,y)$. Correspondingly, we resolve the three-dimensional position vector as $\textbf{R}=(\Rin, z)$. For further convenience, we introduce the in-plane magnetic field component, $\Bin=(B_x,B_y)$. If the magnetic field is constant, which we assume, it is useful to choose the following vector potential:
\be
\textbf{A} = (z-z_0) \Bin \times \z +  \frac{1}{2} \Bzvec \times \textbf{r},
\ee
corresponding to the in-plane and out-of-plane magnetic field components, respectively.
Dropping the zero $z$ component from these two vectors, we introduce
\be
\Ain = (z-z_0) (B_y, -B_x),
\label{eq:Ain}
\ee
with the constant $z_0$ specified below, and
\be
\Az = \frac{1}{2} \Bz (-y, x).
\ee
Both $\Ain$ and $\Az$ are in-plane vectors.
Finally, we write the momentum as $\textbf{P} \equiv (\Pin +e\, \Ain,p_z)$, introducing 
\begin{subequations}
\label{eq:P}
\begin{eqnarray}
\Pin
&=&  -i\hbar(\partial_x,\partial_y) + e\,\Az \nonumber
\\&=& -i\hbar(\partial_x,\partial_y) + \frac{e \Bz}{2}(-y, x), \label{eq:Pin} \\
p_z &=& -i\hbar \partial_z, \label{eq:Pz}
\end{eqnarray}
\end{subequations}
as the in-plane and out-of-plane kinetic-momentum operators,\cite{smrcka1994:JPCM} respectively. The former includes the effects of the perpendicular component of the magnetic field, which is the only way the orbital effects of the magnetic field enter in the quasi-two-dimensional limit.

\subsection{Mixing due to orbital effects of  in-plane field}

With the above definitions, the Hamiltonian in Eq.~\eqref{eq:H zeroth order} can be written as
\be
H=H_{\rm 2D} + H_z + H^\prime_B.
\ee
The first term contains only in-plane coordinates,
\be
H_{\rm 2D} = \frac{\Pin^2}{2m} +V_{\rm 2D}(\Rin),
\label{eq:H2D}
\ee
and the second one only the perpendicular coordinate,
\be
H_z=\frac{p_z^2}{2m} + v(z).
\label{eq:Hz}
\ee
The two sets of coordinates are coupled by the in-plane magnetic field,
\be
H^\prime_B = \frac{e}{m}\Ain \cdot \Pin + \frac{e^2}{2m} \Ain^2 \equiv H^\prime_1 + H^\prime_2,
\label{eq:Hmix}
\ee
where we denoted separately the term linear and quadratic in the in-plane magnetic-field components as $H^\prime_1$ and $H^\prime_2$, respectively. Before continuing, it is useful to note the following identity,
\be
H^\prime_1= \left[ \frac{e}{i\hbar} \Ain \cdot \Rin ,H_{\rm 2D}  \right],
\label{eq:Lm1}
\ee
which can also be written as
\be
\frac{e}{i\hbar} \Ain \cdot \Rin = L_{\rm 2D}^{-1} (H^\prime_1),
\ee
using $L_{\rm 2D}(X) \equiv [X,H_{\rm 2D}]$ as the definition of the Liouville operator $L_{\rm 2D}$ corresponding to $H_{\rm 2D}$, the in-plane Hamiltonian.

\subsection{Symmetries of the confinement potentials}

In the following, we derive results in a general form which does not refer to the specifics of the confinement potentials. However, it is useful to consider certain typical cases. Concerning the dot, we take an anisotropic harmonic confinement,
\be
V_{\rm 2D}(\Rin)
= \frac{\hbar^2}{2m} \left( \frac{ x_d^2}{l_x^4} + \frac{y_d^2}{l_y^4} \right),
\label{eq:V2D}
\ee
parameterized by two confinement lengths, $l_x$ and $l_y$, or, alternatively, the associated energies $\hbar \omega_{x,y} = \hbar^2/m l_{x,y}^2$. If the two are equal, the quantum dot has rotational symmetry in the plane and the eigenstates of $H_{\rm 2D}$ form the Fock-Darwin spectrum. If $l_x \neq l_y$, the dot has two reflection axes $\x_d, \y_d$ which are in general misaligned from the crystallographic axes $\x, \y$ by angle $\delta$. Apart from symmetry, the in-plane excitation energies are of interest. We denote them by $E^*_x$ and $E^*_y$. For the harmonic confinement at zero magnetic field, $E^*_{x,y}=\hbar \omega_{x,y}$, and we denote the energy of this order as $\hbar \omega$. A finite perpendicular magnetic field will change the value of this energy compared to its $\Bz=0$ value,\cite{rebane1969,davies1985,schuh1985} but we will not consider cases where this effect would be substantial.

Concerning the heterostructure confinement, we will include three typical choices. The first is a harmonic confinement,
\be
v_{H}(z)
= \frac{\hbar^2}{2m l_z^4} z^2.
\label{eq:VH}
\ee
It represents structures with $\z$-reflection symmetry. Although it might be realized by modulating the heterostructure composition,\cite{salis2001:N} rather than being microscopically faithful, its advantage is that it results in an analytically solvable model (see Appendix \ref{app:exact}). The second one is a rectangular confinement,
\be
v_{R}(z) = \left\{
\begin{tabular}{ll}
$0$, &if $z \in \langle -l_z/2, l_z/2 \rangle$,\\
$V_0$, &if $z \notin \langle -l_z/2, l_z/2 \rangle$.\\
\end{tabular}
\right.
\label{eq:VR}
\ee
It is a more realistic microscopic description than Eq.~\eqref{eq:VH} for a symmetric quantum well. Here, $V_0$ is the offset of the conduction bands of the two materials defining the quantum well and $l_z$ is its nominal width. The third choice  is a triangular potential,
\be
v_{T}(z) = \left\{
\begin{tabular}{ll}
$V_0$, &if $z<0$,\\
$e E_{\textrm{ext}} z$, &if $z>0$,\\
\end{tabular}
\right.
\label{eq:VT}
\ee
which represents asymmetric cases, for example, a single interface heterostructure with the band offset $V_0$, and the interface electric field $E_{\textrm{ext}}$, which typically arises from a remote doping layer. With this choice, the eigenfunctions can be expressed by Airy functions. They are given, together with several matrix elements which will be needed below, in Appendix \ref{app:matele}. Unlike for previous choices, there is no nominal length $l_z$ in Eq.~\eqref{eq:VT}. It is, however, useful to define it  by $eE_{\textrm{ext}} \equiv \hbar^2/2ml_z^3$ [see Eq.~\eqref{eq:lzapp} in Appendix \ref{app:matele}].

To allow for comparison of confinements with different shapes, we use the following common notation. The ``nominal'' length $l_z$ is considered as a parameter defining the confinement, which is fixed by the fabrication, and therefore does not change (for example, upon the application of the magnetic field). This fixed length defines an associated energy scale $\hbar \omega_z = \hbar^2/ m l_z^2$. These nominal parameters are usually not directly accessible. Instead, spectroscopy can reveal the excitation energies. We denote by $E_z^*$ the energy difference of the lowest two subbands, the subband excitation energy, and we associate the length $l_z^*$ to it by $E^*_z \equiv \hbar^2 / m l_z^{* 2}$. These quantities will change with the magnetic field. Also, at zero magnetic field, even though for the harmonic potential $l_z^*=l_z$, these two lengths differ by factors of order 1 for the other two potentials (see Appendix \ref{app:matele}).

\begin{figure}
\includegraphics[width=1\columnwidth]{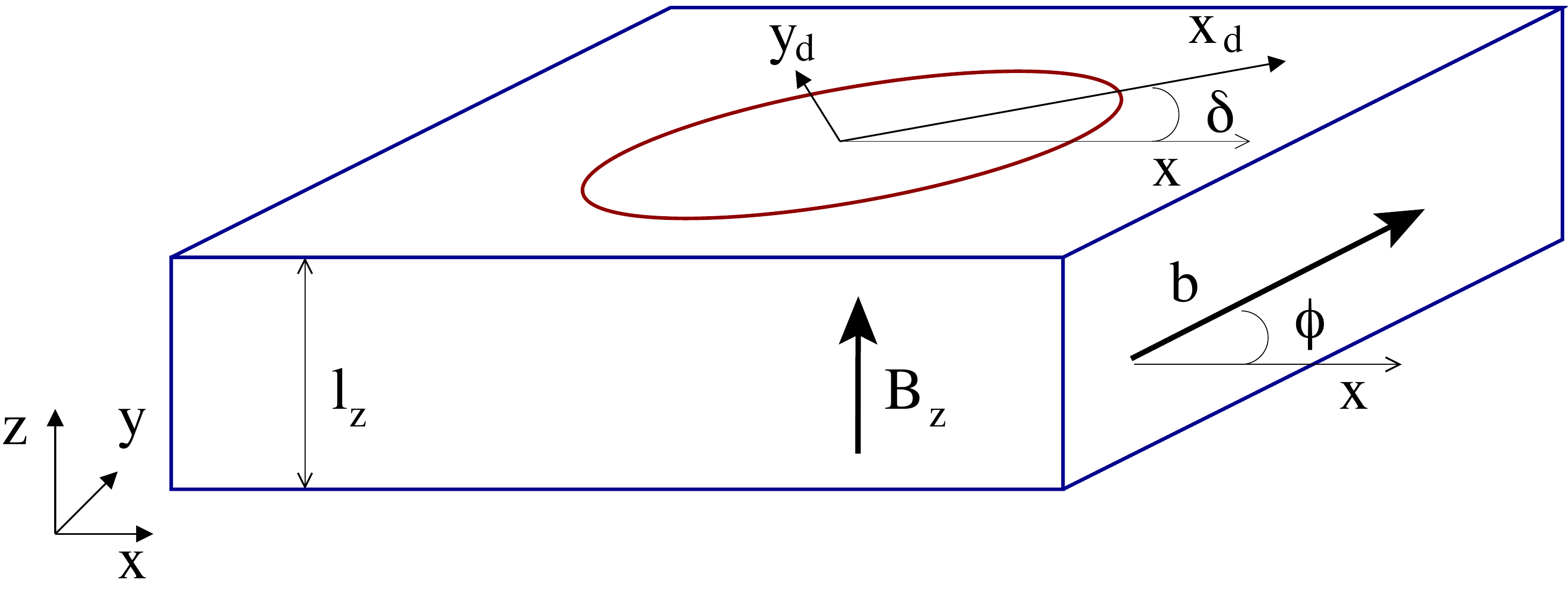}
\caption{\label{fig:sketch}
Geometry of the setup. The quantum dot is defined at the heterostructure interface which is perpendicular to the $\z=[001]$ axis, and has nominal width $l_z$. In the interface plane, the main axis of the quantum dot confinement $\x_d$ makes angle $\delta$ with the crystallographic axis $\x=[100]$. The magnetic field has an out-of-plane component $\Bz$, and the in-plane component $\Bin$, the latter making angle $\phi$ with $\x$.}
\end{figure}

The ratio of the in-plane and perpendicular confinement energies, $\eta=\hbar \omega / \hbar \omega_z$, quantifies how much the dot deviates from the idealized, purely quasi-two-dimensional case (for which $\eta=0$). We call this parameter the aspect ratio. As we are interested in quantum dots that are at least approximately two-dimensional, we will treat this ratio as a small parameter. The importance of the orbital effects of the in-plane field, which are the content of this work, are proportional to $\eta$.  A typical value in gated dots is $\eta =1/10$, or smaller.
The geometry of the structure is summarized in Fig.~\ref{fig:sketch}.

\section{Gauge invariant perturbation theory}

We now perform a perturbative calculation of the orbital effects of the in-plane magnetic field. We will use the second-order degenerate perturbation theory and pay special attention to the gauge invariance.

\subsection{Basis}

The orbital effects of the in-plane field arise through $H^\prime_B$ [Eq.~\eqref{eq:Hmix}]. This term is treated as a perturbation, so that the rest of the Hamiltonian defines the basis. It spans a linear space defined by basis states
\be
|\alpha i\rangle = |\alpha\rangle \otimes |i\rangle,
\label{eq:basis 3D}
\ee
with the corresponding energy $E_{\alpha i}$. The basis state is a tensor product
of an eigenstate of the heterostructure Hamiltonian $H_z$ [Eq.~\eqref{eq:Hz}], with the corresponding wave function
\be
\psi_\alpha(z) = \langle z | \alpha \rangle,
\label{eq:basis 1D}
\ee
and an eigenstate of the 2D quantum dot Hamiltonian $H_{\rm 2D}$ [Eq.~\eqref{eq:H2D}], with the corresponding wave function
\be
\Psi_i(\Rin) = \langle \Rin |i \rangle.
\label{eq:basis 2D}
\ee
We will use the Greek and Roman letters, respectively, as labels of the two sets. Also, we use the standard nomenclature and call a subset of Eq.~\eqref{eq:basis 3D} with a fixed $\alpha$ a subband.

We note that it might be tempting to include $H^\prime_2$, which is a function of $z$ only, into Eq.~\eqref{eq:Hz}. Especially for the harmonic potential, it is simple to find the spectrum of such a redefined Hamiltonian $H_z$ analytically, and find immediately, for example, the expected diamagnetic energy shifts of the subbands. This choice would, however, make the basis gauge dependent, and this not only for the wave functions but also the energies. The gauge invariance of the total Hamiltonian eigenenergies would then be reinstated order by order from the effects of $H^\prime_1$. We therefore find it natural to keep $H^\prime_2$ as a part of the perturbation, making the gauge invariance much more transparent, as we show shortly below.

\subsection{Second order perturbation theory \label{subsec:Lowdin}}

Once the basis has been set, we are ready to evaluate the effects of $H^\prime_B$. We use the degenerate perturbation theory of Ref.~\citenum{birpikus}, which derives an effective Hamiltonian describing a quasi-degenerate subspace.\footnote{The method is known under several names. Our Eq.~\eqref{eq:Lowdin} is taken from Ref.~\citenum{birpikus} [see formula (15.46) on p.~138 therein], which calls it the ``method of successive transformations [of the degenerate perturbation theory]''. Reference \citenum{loehr} calls it a ``method of infinitesimal basis transformations'' (see p.~11 therein), and points out a difference to the ``L\"owdin'' perturbation theory (see Appendix A  p.~233 therein): while both of these are perturbation theories for the effective Hamiltonian, they relate similarly as the Rayleigh-Schroedinger to the Brillouin-Wigner perturbation theory. Namely, the former results in a linear eigenvalue equation with an involved structure of the higher order terms. In the latter, it is simple to generate higher order terms in the perturbation expansion, on the expense of getting a non-linear equation with the unknown energy in the denominators.}
For us, this subspace is the subband $\alpha$. Up to the second order in the in-plane magnetic field, the matrix elements of the effective Hamiltonian for the $\alpha$th subband are
\be
\begin{split}
H^{(\alpha)}_{i j} & =  \langle \alpha i | H^\prime_1+H^\prime_2 | \alpha j\rangle +
 \frac{1}{2} {\sum_{\beta k}}^\prime
 \langle \alpha i | H^\prime_1 | \beta k\rangle \\
&
\times  \langle \beta k | H^\prime_1 | \alpha j\rangle  \left( \frac{1}{E_{\alpha i}-E_{\beta k}}
+ \frac{1}{E_{\alpha j}-E_{\beta k}} \right).
\end{split}
\label{eq:Lowdin}
\ee
The sum runs over all values of the indices $\beta$ and $k$ except the following two pairs: $(\beta k) \neq (\alpha i)$ and $(\beta k) \neq (\alpha j)$.

We now split the sum over the subband index $\beta$ to the term $\beta=\alpha$ and the rest, $\beta\neq \alpha$. Adding the former to the first term of Eq.~\eqref{eq:Lowdin} gives, with the help of the identity in Eq.~\eqref{eq:Lm1}, the following operator:
\be
H^{(\alpha)}_\textrm{intra} = \ave{H^\prime_1}{\alpha} + \ave{H^\prime_2}{\alpha} + \frac{1}{2} \left[ \ave{H^\prime_1}{\alpha},\ave{L_{\rm 2D}^{-1}(H^\prime_1)}{\alpha}  \right].
\label{eq:intra}
\ee
It contains terms with the $z$-dependent operators averaged over the given subband profile, $\ave{X}{\alpha} \equiv \langle \alpha | X |\alpha \rangle$. The first term in the previous equation is
\be
\ave{H^\prime_1}{\alpha} = \frac{e}{m}\ave{\Ain}{\alpha} \cdot \Pin.
\ee
This term can be added to Eq.~\eqref{eq:H2D} and removed by a convenient gauge choice for the vector potential.
Specifically, choosing $z_0 = \ave{z}{\alpha}$, it becomes zero. Note, however, that in general the gauge removal of this term can be done only within a single subband. This is natural, since if wave functions of two subbands differ in their center of mass along the $z$ coordinate [which is the case, for example, for the triangular potential in Eq.~\eqref{eq:VT}], the in-plane field has to result in phases upon intersubband transitions. If these phases are of relevance,\cite{choi1988:PRB,yang1997:PRL} $\ave{H^\prime_1}{\alpha}$ should be included in Eq.~\eqref{eq:H2D} and kept track of explicitly (in another words, a single choice for $z_0$ has to be made for all subbands). On the other hand, in a symmetric heterostructure potential, all subbands have the same center of mass and a single choice removes $\ave{H^\prime_1}{\alpha}$ for all subbands. For the symmetric confinements given in Eqs.~\eqref{eq:VH} and \eqref{eq:VR}, this would be the choice $z_0=0$.

We also note that such gauge removal is not possible for $H^\prime_1$ itself, where $z$ is still an operator. The difference is illustrated by the following. The remaining two terms from Eq.~\eqref{eq:intra} produce the subband diamagnetic shift,\cite{stern1967:PR}
\be
E_\textrm{dia}^{(\alpha)} =  \frac{e^2}{2m} \Bin^2 \, \textrm{var}_\alpha (z).
\label{eq:Ediag}
\ee
Here, the variance is defined by
\be
\textrm{var}_\alpha (z) = \ave{(z-z_0)^2}{\alpha}-\left(\ave{(z-z_0)}{\alpha}\right)^2,
\ee
and is clearly independent on $z_0$, that is, gauge invariant. The second term, required for the expression to be invariant to the choice of $z_0$, comes from the $H_1^\prime$ term.

\subsection{Recipe}

We summarize the above in the following recipe. Interested in the in-plane field effects on the lowest subband $\alpha$, the choice $z_0 = \ave{z}{\alpha}$ reduces the effective Hamiltonian for this subband to the sum of three terms. A purely 2D quantum dot Hamiltonian, Eq.~\eqref{eq:H2D}, the diamagnetic shift (an overall constant) $E_\textrm{dia}^{(\alpha)}$ [Eq.~\eqref{eq:Ediag}], and the following correction:
\be
\begin{split}
\langle i | H^{(\alpha)}_\textrm{inter} | j \rangle = & \frac{1}{2}
{\sum_{\beta \neq \alpha}}
{\sum_{k}}
 \langle \alpha i | H^\prime_1 |\beta k\rangle\langle \beta k | H^\prime_1 | \alpha j\rangle  \\
& \qquad
\times \left( \frac{1}{E_{\alpha i}-E_{\beta k}}
+ \frac{1}{E_{\alpha j}-E_{\beta k}} \right).
\end{split}
\label{eq:Lowdin2}
\ee
The latter is a sum of contributions from all subbands $\beta$ other than $\alpha$, and is expressed through
\be
\langle \alpha | H^\prime_1 | \beta \rangle = \frac{e}{m}
z_{\alpha\beta} \left( \Bin \times \z \right) \cdot \Pin,
\ee
an operator in the in-plane coordinates only. It depends on the dipole matrix elements of the $z$ coordinate
\be
z_{\alpha \beta}= \langle \alpha | z | \beta \rangle,
\ee
 and is therefore also explicitly independent of $z_0$, the choice of the gauge.

\subsection{Small aspect-ratio approximation}

\label{sec:smallaspect}

\begin{figure}
\includegraphics[width=0.9\columnwidth]{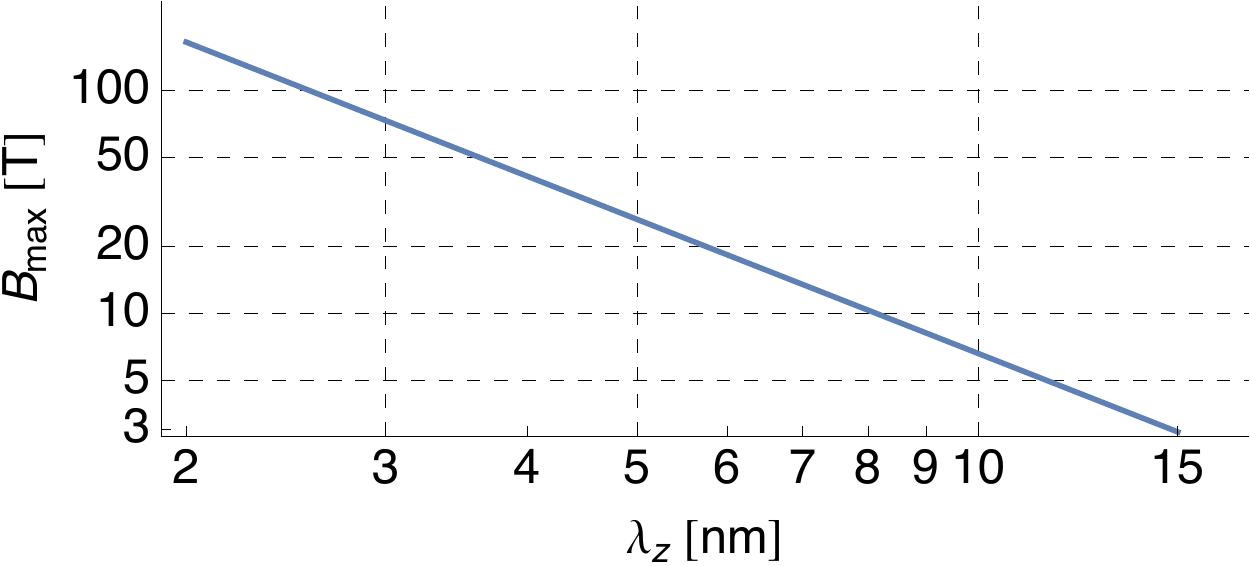}
\caption{\label{fig:Bmax}
The magnitude of the in-plane magnetic field at which the flux $\flux$ [Eq.~\eqref{eq:phi}] reaches unity. For fields smaller, corresponding to $\Phi<1$, the expansion in $H_1^\prime$ is convergent and Eq.~\eqref{eq:Hainter} is the leading correction. For larger magnetic field, this expression is still of use, if it is corrected by the phenomenological replacement given in Eq.~\eqref{eq:replacement}.}
\end{figure}

The expression for the inter-subband correction in Eq.~\eqref{eq:Lowdin2} can be further simplified by using perturbation theory  in the aspect ratio $\eta=\hbar \omega / \hbar \omega_z$. In leading order, neglecting the in-plane excitation energies with respect to the subband excitation energies in the denominators, we get
\be
H^{(\alpha)}_\textrm{inter} = -\flux^2 \, \frac{[\Pin \cdot (\Binunit \times \z ) ]^2}{2m}  + O\left(\eta \right),
\label{eq:Hainter}
\ee
where we denoted $\Binunit$ as the unit vector along the in-plane component of the magnetic field $\Bin$.
Finally, 
\be
\flux = \frac{e}{\hbar} \Binscalar  \lambda_z^2,
\label{eq:phi}
\ee
is the dimensionless flux due to the in-plane magnetic field through the 2DEG effective width $\lambda_z$  squared.\footnote{The dimensionless flux is usually defined using a flux quantum $h/e$, instead of $\hbar/e$ which appears in Eq.~\eqref{eq:phi}. We opt for this choice to prevent factors $2\pi$ appearing either in Eq.~\eqref{eq:Hainter} or \eqref{eq:lambdaz}.}
The latter is defined by\cite{stern1968:PRL}
\be
\lambda_z^{4}= 2{\sum_{\beta \neq \alpha}} \frac{\hbar^2}{m} \frac{|z_{\alpha \beta}|^2}{E_\beta-E_\alpha},
\label{eq:lambdaz}
\ee
as a sum of contributions from all subbands except $\alpha$.

The result in Eq.~\eqref{eq:Hainter} is worth commenting. It states that the dominant effect of the in-plane field is a renormalization of the particle mass along the direction perpendicular to the in-plane component of the magnetic field. In the lowest subband, the particle becomes heavier along this direction. The effect is proportional to $\flux^2$, the second power of the flux due to the in-plane magnetic field through the area defined as the square of the  effective 2DEG-width $\lambda_z$.\footnote{It also means that the kinetic energy is still time-reversal symmetric. One has to go to the next order in the perturbation theory to obtain an asymmetric term, which has importance, for example, for weak localization effects.\cite{falko2002:PRB}} This flux plays also the role of the small parameter for the perturbation in $H_1^\prime$, and the condition $\flux \ll 1$ is the condition for Eq.~\eqref{eq:Lowdin2} to be the dominant term. Figure \ref{fig:Bmax} shows the magnetic field at which the flux becomes one. 
Finally, all the details of the heterostructure confinement are reduced to a single parameter, $\lambda_z$, the effective width of the 2DEG.

\subsection{The effective 2DEG width}

\begin{table}
\begin{tabular}{ccccc}
\hline \hline
confinement\,\, & \multicolumn{2}{c}{effective width}  & lowest \,\,& exc. energy \,\,\\
shape& $\lambda_z/l_z$ & $\lambda_z/l_z^*$ & exc. sub.& $E_z^* / \hbar \omega_z$ \\
\hline
harmonic & 1  & 1 & 100\% & 1\\
rectangular & 0.257  & 0.99 & 99.9\% & 14.8\\
triangular & 1.01 & 0.943  & 94.3\% & 0.875\\
\hline \hline
\end{tabular}
\caption{\label{tab:lambdaz}
The parameters related to the 2DEG effective width for various confinements. The confinement shape is given in the first column. The second column gives the effective width $\lambda_z$, Eq.~\eqref{eq:lambdaz}, in units of the nominal width $l_z$, defined for each potential shape individually [see Eqs.~\eqref{eq:VH}--\eqref{eq:VT}]. This equation is therefore to be used if the microscopic parameters of the confinement are known. The third column gives the effective width in the units of a length scale derived from the subband excitation energy $E_z^* \equiv \hbar^2/ml_z^{*2}$ and is therefore useful if the latter is known. The fourth column gives the relative weight of the lowest excited subband contribution to the effective width. The last column gives the energy distance to this subband in units of $\hbar \omega_z$.  The results for the triangular and rectangular potentials are given in  the limit $V_0 \to \infty$, and would change  very little upon using a typical value of $V_0$ in GaAs, such as 300 meV, instead.}
\end{table}

We calculate $\lambda_z$ in Appendix \ref{app:matele} for the three confinement choices as a function of their respective natural parameters, and summarize the results in Table \ref{tab:lambdaz}. From the latter one can see that for the choices that we considered, Eqs.~\eqref{eq:VH}--\eqref{eq:VT}, there is little variation among different confinements, if the effective length is related to the subband excitation energy $E_z^*$ or, equivalently, $l_z^*$. Within the typical precision of Eq.~\eqref{eq:Hainter}, one can therefore set
\be
\lambda_z \approx \frac{\hbar}{\sqrt{m E_z^*}},
\label{eq:lambdazapprox}
\ee
irrespective of the perpendicular confinement shape.

\begin{figure}
\includegraphics[width=1\columnwidth]{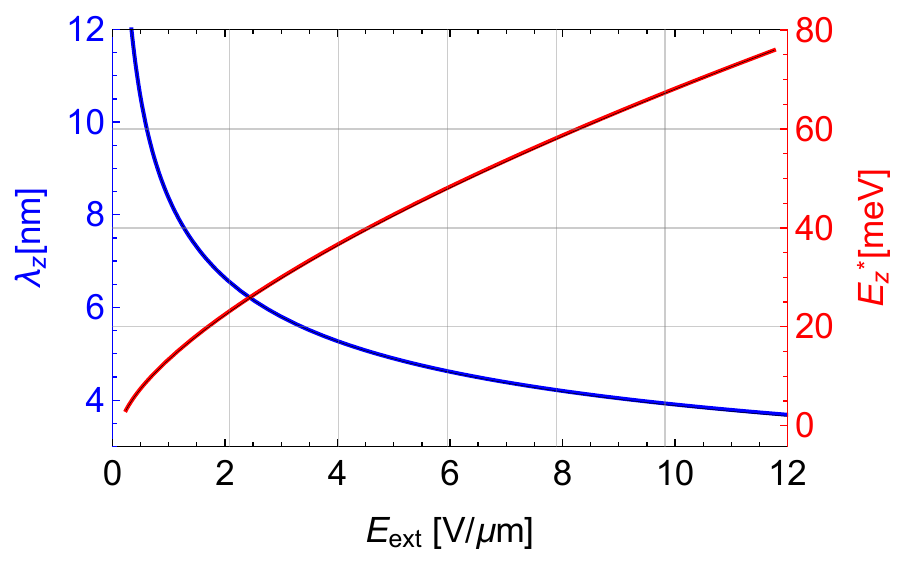}
\caption{\label{fig:lambdaz}
The 2DEG effective width $\lambda_z$ (left $y$ axis) and the subband excitation energy $E_z^*$ (right $y$ axis) for the triangular confinement potential, Eq.~\eqref{eq:VT}, as a function of the interface electric field. The thick colored lines are obtained from numerical solutions for the confinement with a typical value for the conduction band offset $V_0=300$ meV, while the thin black lines are analytical results in the limit $V_0 \to \infty$. The two sets of lines are indistinguishable on the figure resolution, illustrating that one can safely use the infinite offset approximation when evaluating the two quantities of interest.
}
\end{figure}

For completeness, for each confinement we now express it in its natural parameters given in Eqs.~\eqref{eq:VH}--\eqref{eq:VT}. For the harmonic confinement, the length $l_z$ is defined through the potential curvature, which results in the exact relations, $\lambda_z=l_z=l_z^*$ and $E^*_z=\hbar \omega_z$. Only the lowest excited subband contributes in Eq.~\eqref{eq:lambdaz}; the dipole matrix elements for all other subbands are zero. Next, the rectangular potential can also be solved analytically in the limit $V_0 \to \infty$, resulting in the expressions given in Table \ref{tab:lambdaz}. We have checked in Appendix \ref{app:matele} that this limit is a very good approximation for realistic values of the offset $V_0$. Finally, the triangular potential is the only one for which the contributions from the higher subbands are sizable, though still small compared to the lowest one. We conclude that concerning the effective length, the heterostructure shape is of little relevance, determined mostly by the subband excitation energy, and contributed to mostly by the lowest excited subband. Choosing the triangular potential, we illustrate the relations between the effective length, the microscopic parameters (being here the interface electric field and the conduction band offset), and the subband excitation energy in Fig.~\ref{fig:lambdaz}. To conclude this section, Eqs.~\eqref{eq:Hainter}--\eqref{eq:lambdazapprox} allow one to grasp the leading orbital effects of an in-plane field in a very simple way.

\section{Effects on spectrum: fingerprints of quantum dot orbitals}

The orbital effects of the in-plane field can be exploited as a tool to characterize the quantum dot. To demonstrate the principle, we first assume that the magnetic field is purely in-plane, $\Bz=0$, and that the corresponding flux is small, $\Phi \ll 1$. The effects of $H_\textrm{inter}^{(\alpha)}$ [Eq.~\eqref{eq:Hainter}] can then be treated perturbatively. The eigenstates of the unperturbed subband-Hamiltonian, $H_{\rm 2D}$, with the anisotropic harmonic confinement given in Eq.~\eqref{eq:V2D}, can be labeled by a pair of non-negative integers $n_x$, $n_y$. They correspond to the quantum numbers of two harmonic oscillators with energies $\hbar \omega_x$, and $\hbar \omega_y$, respectively. The expectation value of $H_\textrm{inter}^{(\alpha)}$ in such an eigenstate is
\be
\begin{split}
\delta E_{n_x,n_y} & = - \frac{\Phi^2}{2} \Bigg[ \hbar \omega_x \sin^2(\delta-\phi)  \left( n_x+\frac{1}{2} \right)\\
&  \qquad +  \hbar \omega_y  \sin^2(\delta+\pi/2-\phi) \left( n_y+\frac{1}{2} \right) \Bigg].
\end{split}
\label{eq:dEnxny}
\ee
As an example, the ground state energy correction is
\be
\delta E_{0,0} = - \frac{\Phi^2}{4} \left( \frac{\hbar \omega_x +\hbar \omega_y}{2} - \frac{\hbar \omega_x -\hbar \omega_y}{2} \cos (2\delta-2\phi) \right).
\label{eq:dE00}
\ee
The correction oscillates upon changing the in-plane field direction with period $\pi$. The magnitude of the variation reveals the anisotropy of the confinement potential, as the difference of the two characteristic energies $\hbar \omega_x -\hbar \omega_y$. The energy minimum corresponds to a magnetic field being aligned along the soft confinement axis.

Alternatively, one can look at the changes of the two excitation energies (that is, the energy offsets of the two lowest excited states with respect to the ground state),
\begin{subequations}
\label{eq:dEstar}
\begin{eqnarray}
\delta E^*_x &=& -  \frac{\Phi^2}{2} \hbar \omega_x \sin^2(\delta_x-\phi), \\
\delta E^*_y &=& -  \frac{\Phi^2}{2} \hbar \omega_y \sin^2(\delta_y-\phi).
\end{eqnarray}
\end{subequations}
The excitation energy for a given orbital also oscillates with the same period $\pi$, reaching its maximum when the in-plane magnetic field is aligned with the corresponding ``excitation axis.'' Here, it is $\x_d$, with $\delta_x\equiv \delta$, and $\y_d$, with $\delta_y\equiv \delta+\pi/2$, for the two orbitals, respectively.

We note that the subband Hamiltonian $H_{\rm 2D}+H^{(\alpha)}_\textrm{inter}$, with the second term approximated by Eq.~\eqref{eq:Hainter} can be diagonalized analytically without any further approximations. However, the full formulas give little insight, and we give them only in Appendix \ref{app:exact}. One might be interested in the limit where the dot is so close to being circularly symmetric that  $H_{\textrm{inter}}^{(\alpha)}$ is larger than the difference $\hbar \omega_x-\hbar \omega_y$. In this, nearly-degenerate, case we need to go beyond the non-degenerate perturbation theory used in deriving Eq.~\eqref{eq:dEnxny}. We instead get, in this limit and again for $\Bz=0$, the renormalization of the two excitation energies as
\begin{subequations}
\label{eq:dEstar2}
\begin{eqnarray}
\delta E^*_1 &=& \hbar \omega_+ - \hbar \omega_- \cos(2\delta-2\phi) ,\\
\delta E^*_2 &=& \sqrt{1-\Phi^2} \left[ \hbar \omega_+ + \hbar \omega_- \cos(2\delta-2\phi) \right],
\end{eqnarray}
\end{subequations}
where $\hbar \omega_\pm = (\hbar \omega_x \pm \hbar \omega_y)/2$. The magnitude of the oscillation is proportional to the potential anisotropy, $\hbar \omega_-$, and disappears for a circularly symmetric dot, as expected.

Additional useful information about the quantum dot can be extracted from the  dependence of the energy corrections on the in-plane magnetic field magnitude. Namely, it follows from Eqs.~\eqref{eq:phi} and \eqref{eq:dEstar} that
\be
\lambda_z^4 = -\left. \frac{1}{\sin^2(\delta_i-\phi)} \frac{\hbar^2}{e^2}\frac{1}{E^*_i} \frac{\partial^2 E^*_i}{\partial \Binscalar ^2}\right|_{\Binscalar =0},
\ee
with $i \in \{x,y\}$. The effective width of the 2DEG can be found from the curvature of the excitation energy as a function of the in-plane magnetic field evaluated at $\Binscalar =0$. The shift is largest if the field is applied along the direction given by $\phi=\delta_i+\pi/2$, where the angle $\delta_i$ denotes the orientation of the excitation axis of the corresponding orbital. We point out that it is important that the dot is empty, so that there are no electron--electron interaction effects. These interaction effects make the extraction of the width from analogous measurements in 2DEGs much more involved.\cite{kunze1987:PRB, smrcka1995:PRB, salis1998:PRB, tutuc2003:PRB, gokmen2008:PRB}

We note that one could in principle also use the diamagnetic shift, Eq.~\eqref{eq:Ediag}, to find the effective 2DEG width. Using the flux variable, the shift is
\be
E_\textrm{dia}^{(\alpha)} = \frac{1}{2} \Phi^2 \hbar \omega_z \textrm{var}_\alpha(l_z z /\lambda_z^2),
\label{eq:Ediag2}
\ee
where the constants $\textrm{var}_\alpha(l_z z /\lambda_z^2)$ are of order one (see Appendix \ref{app:matele}). Therefore, the change is larger, by a factor $1/\eta$, compared to the changes of the in-plane excitation energies. However, the issue with trying to measure directly, for example, the lowest subband shift, is that Eq.~\eqref{eq:Ediag2} gives the ``bare'' shift of the given 2DEG subband. With the chemical potential fixed, such a subband shift would change the 2DEG density, resulting in additional electrostatic contributions. In other words, the bare shift of the band bottom is partially screened by the 2DEG. The actual shift can be anywhere between zero and 100\% of the bare shift,\cite{luo1990:PRB} with the ratio (the screening efficiency) given by the 2DEG capacitances to the gates and the self-capacitance.\cite{luryi1988:APL} If this ratio is not known, the measured shift gives only the upper limit for the bare shift, and thus for $\lambda_z$. This problem does not occur for the excitation energies, where the overall subband shift cancels. One could therefore instead consider the diamagnetic renormalization of the subband excitation energy (the equation is valid for the triangular potential) given by
\be
E^*_\textrm{dia} = E^{(\alpha=2)}_\textrm{dia} - E^{(\alpha=1)}_\textrm{dia} \approx \frac{1}{2} \Phi^2 \hbar \omega_z.
\ee
However, due to its relatively large value, the subband excitation energy is not easily accessible; see Ref.~\citenum{lu2017:SR} for an example of its determination in a transport measurement.

\begin{figure}
\includegraphics[width=0.45\columnwidth]{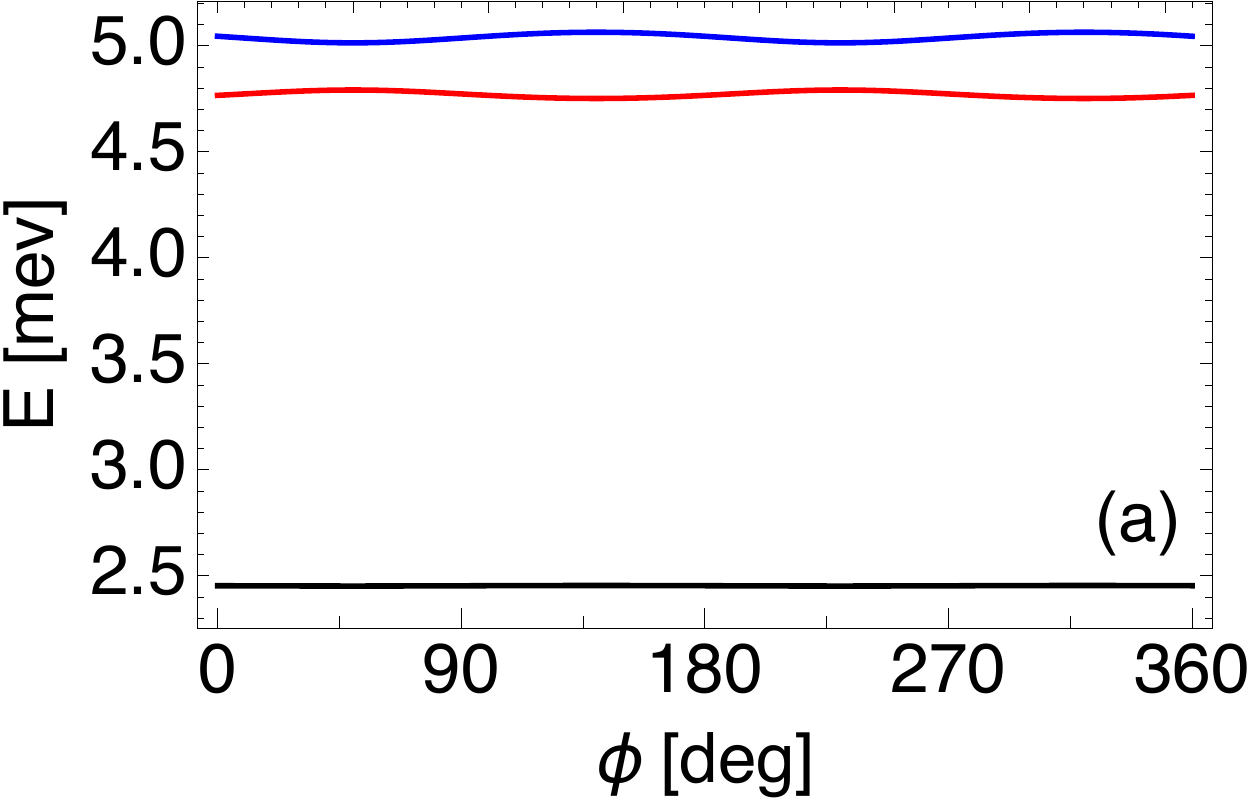}
\includegraphics[width=0.45\columnwidth]{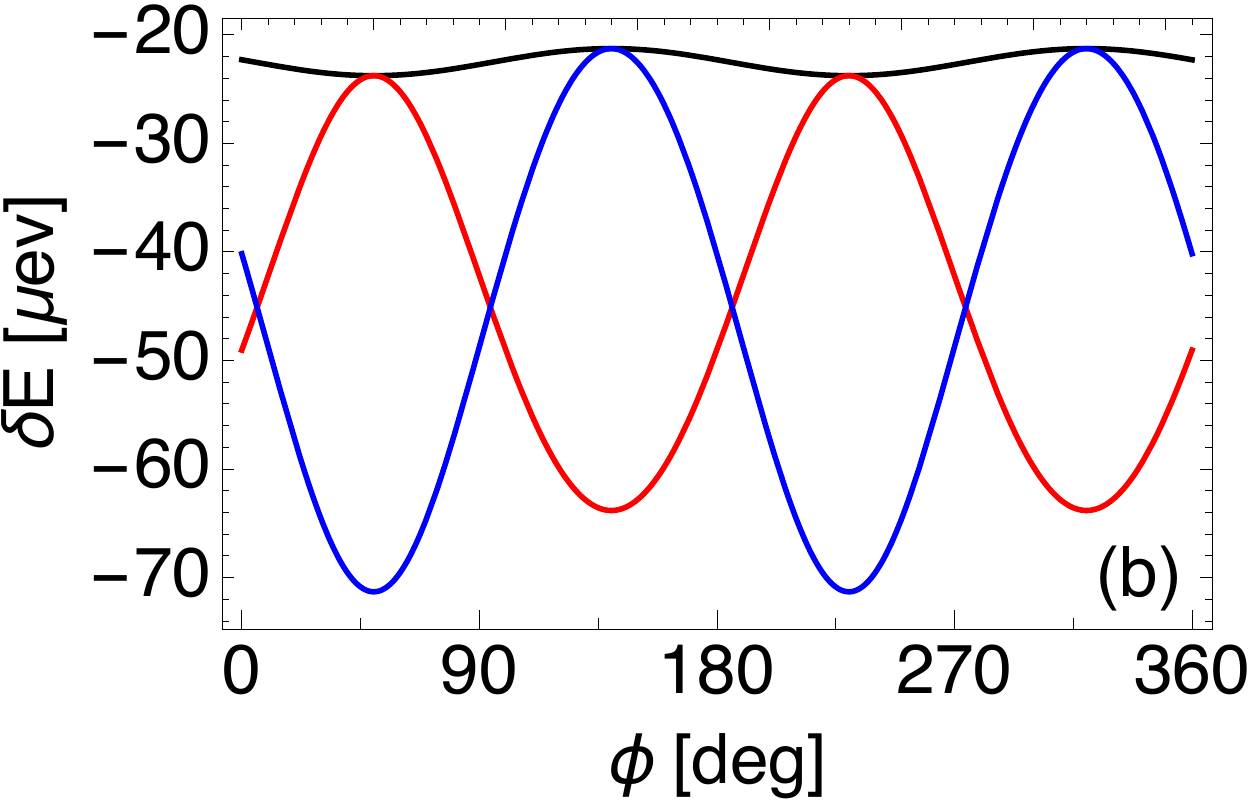}
\includegraphics[width=0.45\columnwidth]{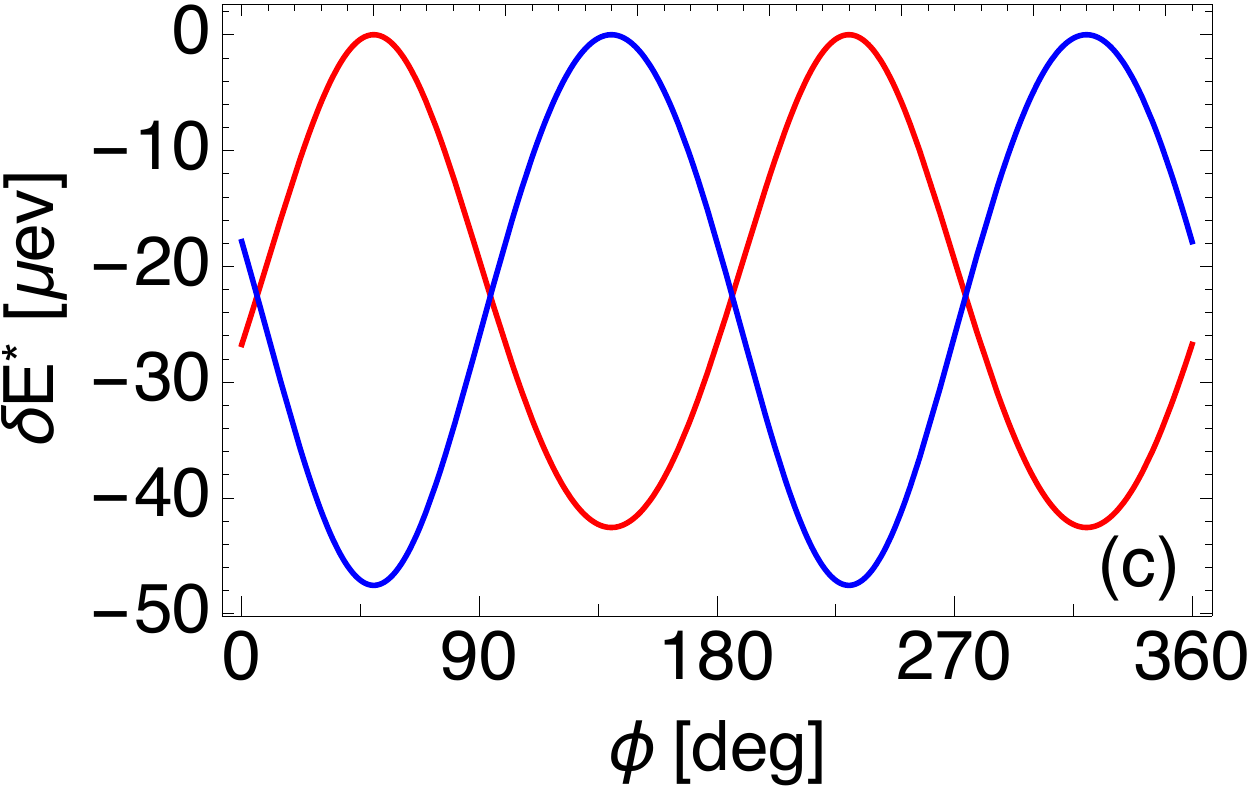}
\includegraphics[width=0.45\columnwidth]{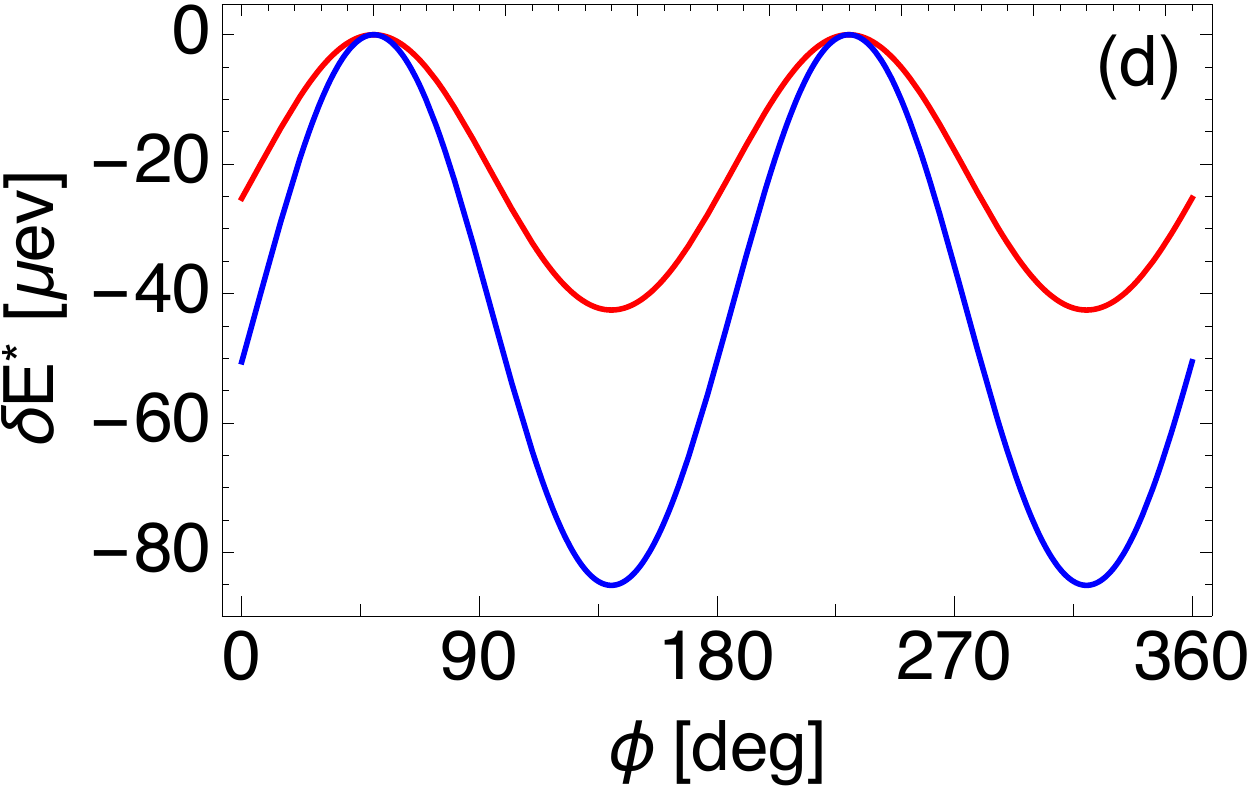}
\caption{\label{fig:illustration}
The spectrum as a function of $\phi$, the direction of the in-plane magnetic field. Unless stated otherwise, we adopt the following parameters (chosen in line with the experiment in Ref.~\citenum{camenzind}, see Fig.~\ref{fig:data} below)
 $\hbar \omega_x=2.34$ meV, $\hbar \omega_y=2.61$ meV, 
$\lambda_z=6.5$ nm,
$m=0.067 m_e$, in-plane field $B=3$ T, and $\delta=50^\circ$. (a) The ground state (black) and the two lowest excited states (red and blue). (b) The corrections to the three eigenenergies (their value at $B=0$ subtracted). (c) The two lowest excitation-energy corrections. (d) The two lowest excitation-energy corrections for a highly anisotropic dot, $\hbar \omega_y=10$ meV.}
\end{figure}

We illustrate these points in Fig.~\ref{fig:illustration}, plotting the energies and their variations as a function of the in-plane field orientation described by the angle $\phi$. We first take a slightly anisotropic dot, with the difference of the two harmonic-oscillator energies approximately 10\% of their average [though still in the limit $|\langle H^{(\alpha)}_\textrm{intra} \rangle |\lesssim \hbar \omega_-$, so that Eq.~\eqref{eq:dEnxny} is valid]. Figure \ref{fig:illustration}(a) shows the energies themselves. The magnitude of the oscillations of the ground state is smaller than that of excited states, as follows from Eqs.~\eqref{eq:dEnxny} and \eqref{eq:dE00}. One can see it more clearly in Fig.~\ref{fig:illustration}(b), which shows only the variations of the energies, subtracting a constant from each of them. The orientation of the  soft and hard axes of the confinement potential is revealed as the angle at which the second and the third energies, respectively, becomes maximal.
A very similar behavior is displayed by the variations of the two excitation energies, plotted in Fig.~\ref{fig:illustration}(c). This behavior can be contrasted with the variations of a much more anisotropic dot, plotted in Fig.~\ref{fig:illustration}(d). Here, the two lowest excited states vary in phase (and their oscillations magnitudes ratio is 2), as they belong to the same orbital. \textit{This characteristic fingerprint can therefore distinguish different types of dots (1D versus 2D), and allows one to determine the spatial orientation of each orbital individually.\cite{yu}}

\section{Accuracy of the perturbative result}

\begin{figure}
\includegraphics[width=0.45\columnwidth]{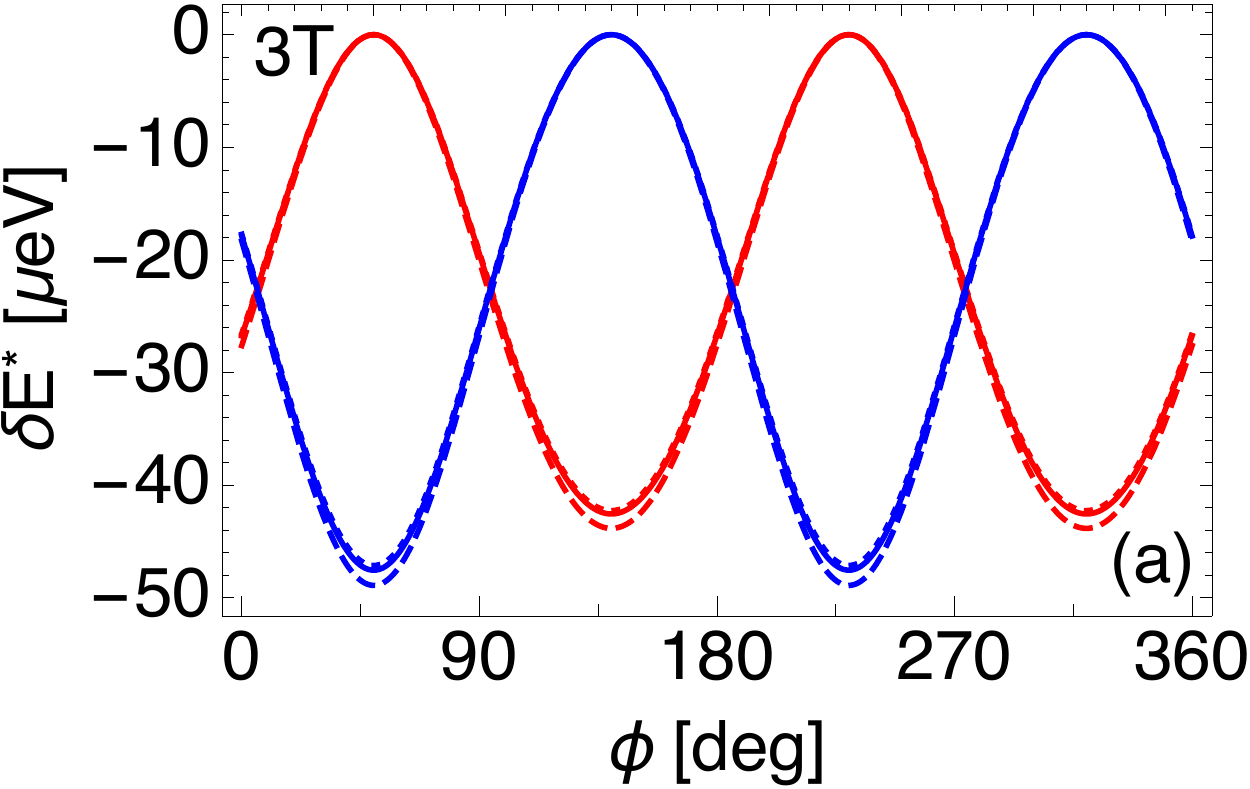}
\includegraphics[width=0.45\columnwidth]{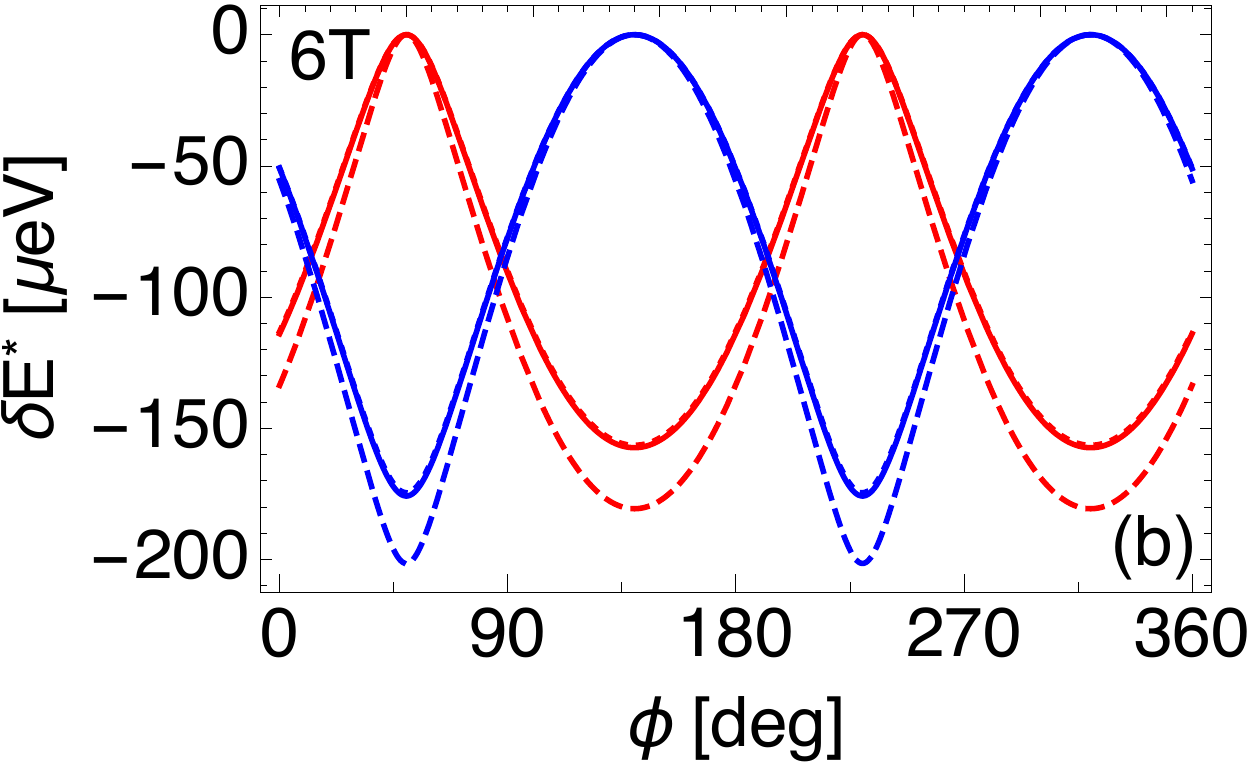}
\includegraphics[width=0.45\columnwidth]{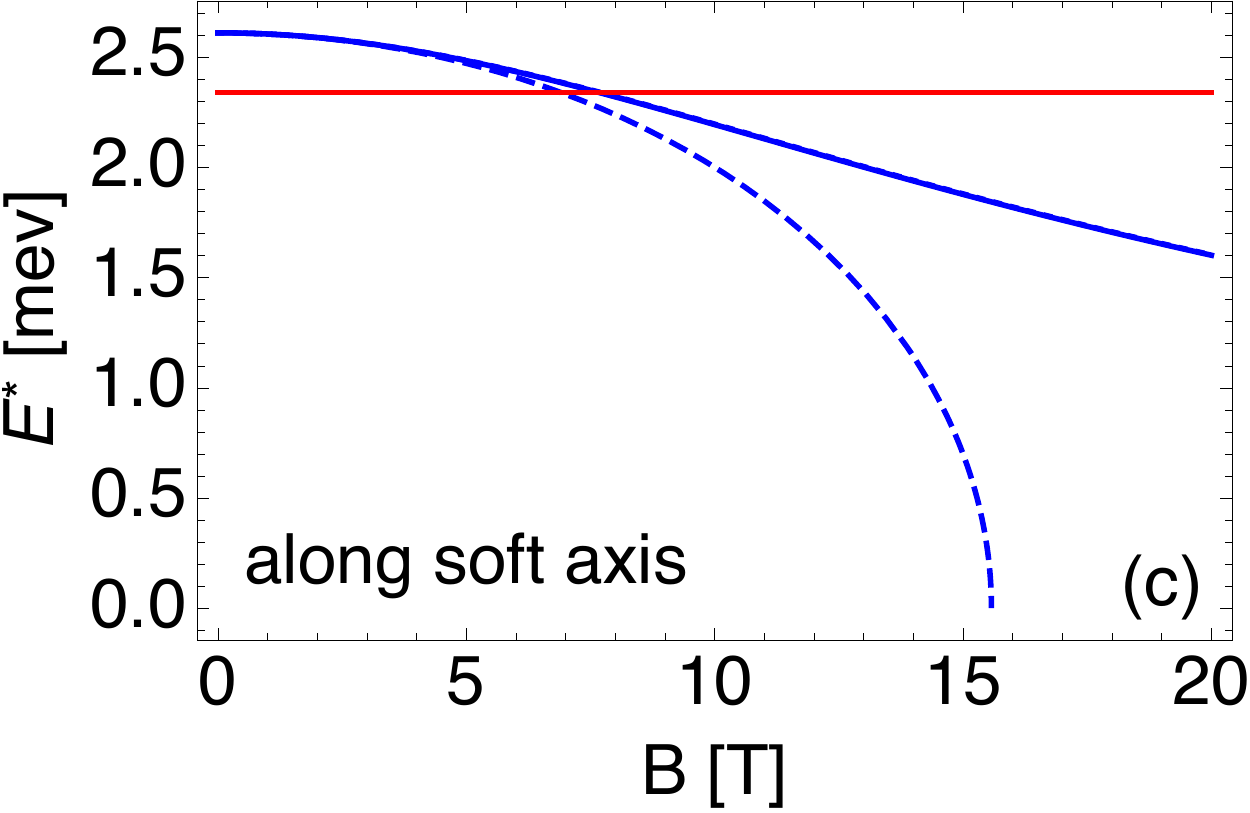}
\includegraphics[width=0.45\columnwidth]{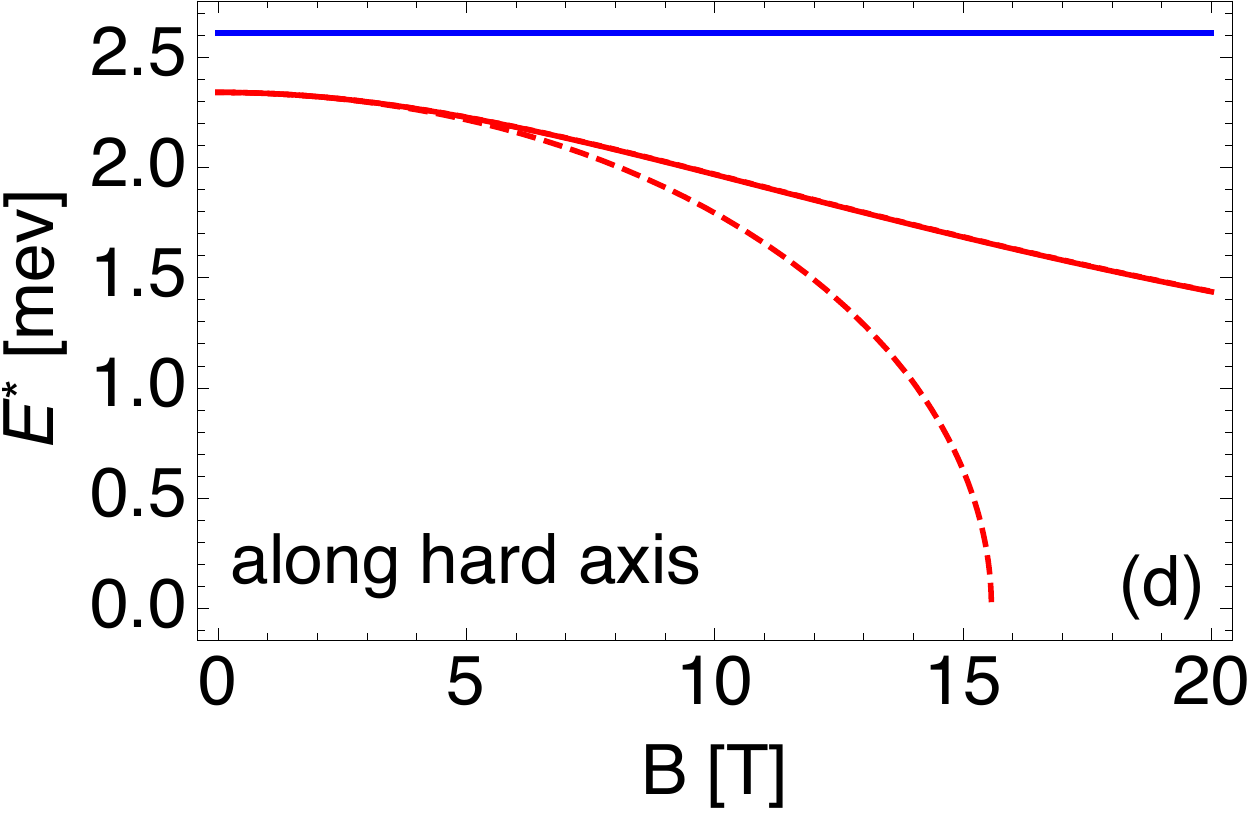}
\caption{\label{fig:comparison}
The excitation energies as a function of the in-plane magnetic-field [(a), (b)] orientation and [(c), (d)] magnitude. The parameters are the same as those in Fig.~\ref{fig:illustration} unless stated otherwise. In (a) and (b), the changes of the two lowest (in red and blue, respectively) excitation energies are plotted for a field of a constant magnitude, as given in the figures, corresponding to (a) $\Phi\approx 0.2$, and (b) $\Phi\approx 0.4$. In (c) and (d), the two lowest excitation energies are plotted as a function of the field magnitude for a fixed direction, (c) $\phi=\delta$ and (d) $\phi=\delta+\pi/2$.
 In all panels, the solid lines are the exact results from the 3D model, the dashed lines are the exact results of the 2D model using Eq.~\eqref{eq:Hainter}, without the replacement in Eq.~\eqref{eq:replacement}. Once this replacement is made, the results of the 2D model become identical to those of the 3D model. See Appendix \ref{app:exact} for details on the models.
}
\end{figure}

\begin{figure}
\includegraphics[width=0.49\columnwidth]{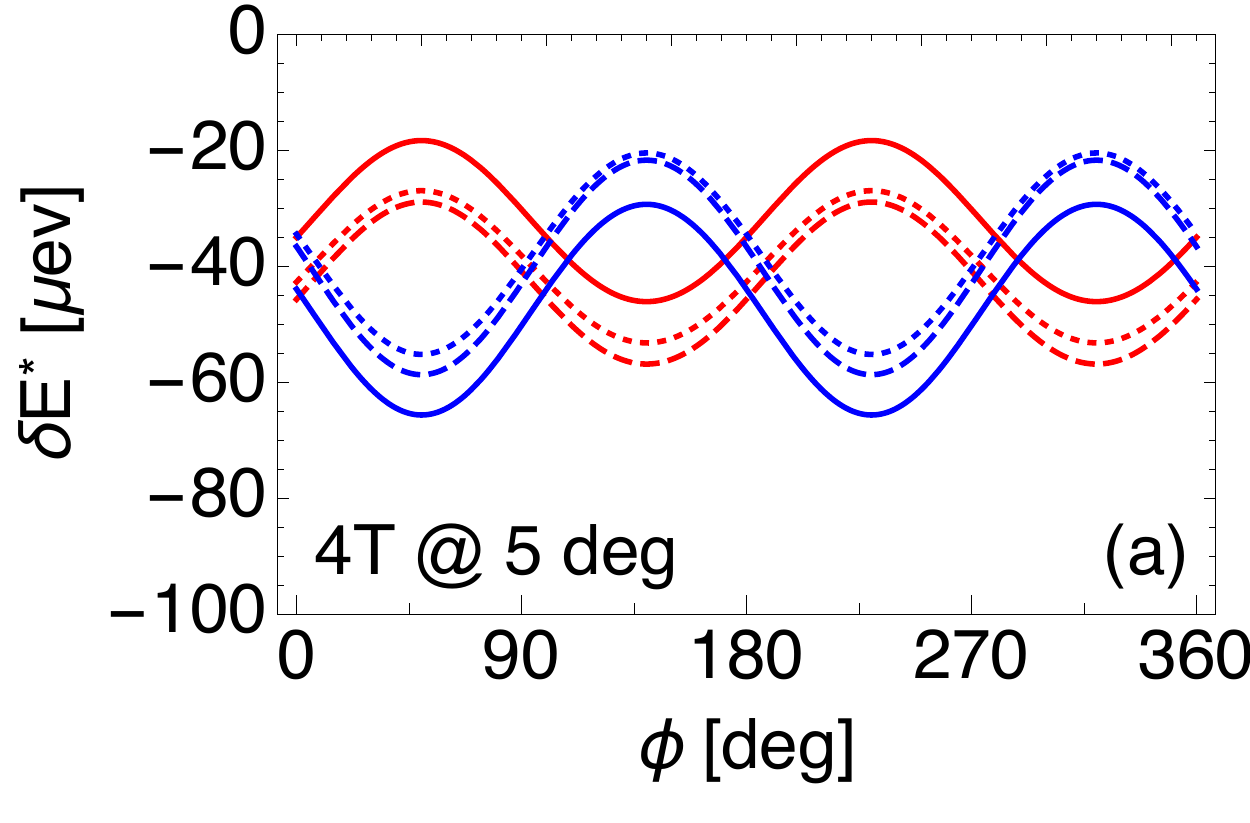}
\includegraphics[width=0.49\columnwidth]{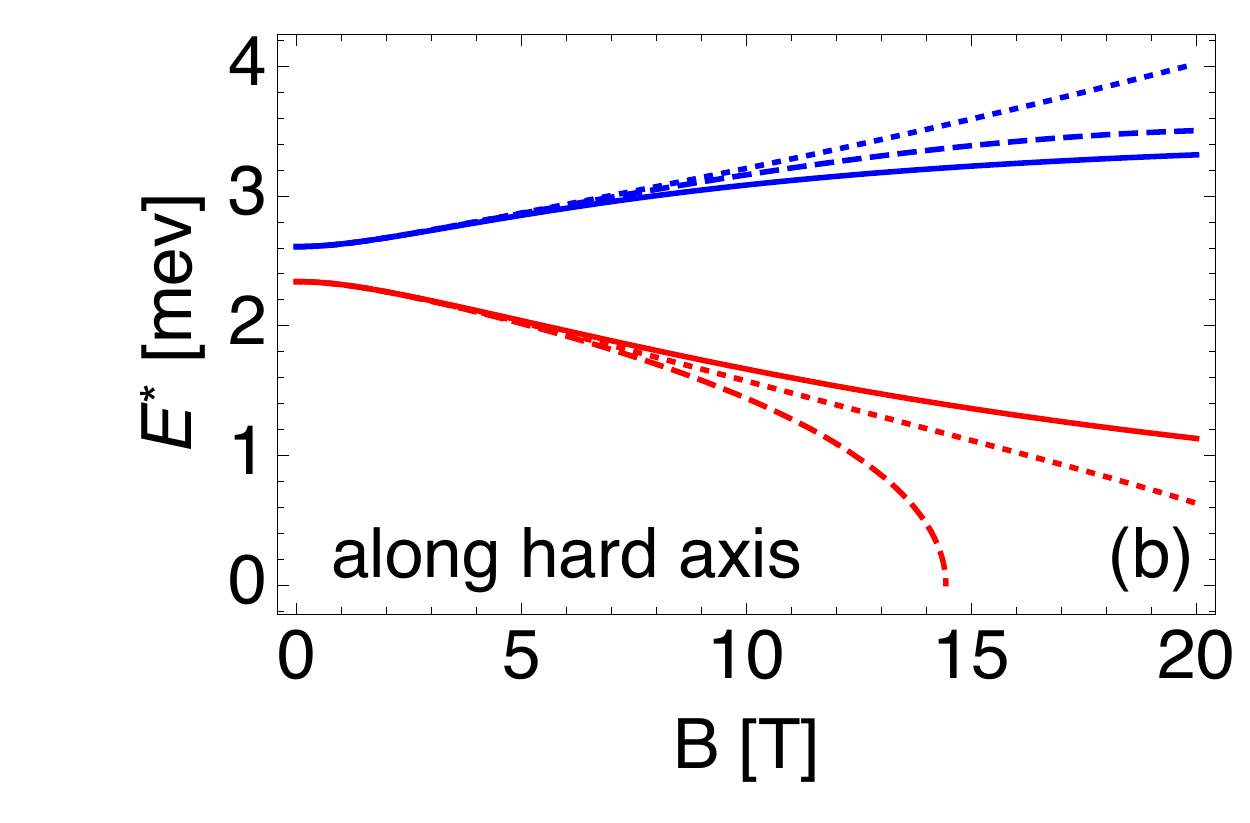}
\caption{\label{fig:compar_add}
The excitation energies as a function of the in-plane magnetic-field (a) orientation and (b) magnitude. The figure is analogous to Fig.~\ref{fig:comparison}, with the magnetic field deflected out of the plane by 5$^\circ$. The solid lines are the exact results from the 3D model, the dashed (dotted) lines are the exact results of the 2D model using Eq.~\eqref{eq:Hainter}, without (with) the replacement in Eq.~\eqref{eq:replacement}. In (b), the divergence of the dashed line towards zero happens at flux $\flux=1$. }
\end{figure}

We now discuss the range of validity and precision of the energy corrections calculated using Eq.~\eqref{eq:Hainter}. To this end, we consider the harmonic heterostructure confinement, Eq.~\eqref{eq:VH}. In this case, the full three-dimensional model has an analytical solution for arbitrary magnetic field (see Appendix \ref{app:exact}), which we can use as a benchmark for the effective two-dimensional model. We obtain the energies of the latter by solving for the spectrum of $H_{\rm 2D} + H^{(\alpha)}_\textrm{intra}$ exactly (see Appendix \ref{app:exact}). We plot the two sets of excitation energies as solid (3D model) and dashed (2D model) lines in Fig.~\ref{fig:comparison}.
Panel (a) shows the directional variation of the energy corrections in an intermediate magnetic field of a few Tesla. Since the chosen parameters correspond to a flux $\Phi\approx 0.2 <1$, the effective 2D model is an excellent approximation to the full 3D model, as expected. Panel (b) shows the energy variations for a larger flux. Even though the directional dependence becomes quite different from a simple sine function, the variations are still correctly reproduced by the 2D model. This model becomes unreliable only when the flux is close to unity.  The reason for this is that for $\Phi=1$, the correction term Eq.~\eqref{eq:Hainter} is so large that the in-plane mass tensor becomes non-positive and the corresponding excitation energy becomes zero [see Figs.~\ref{fig:comparison}(c) and (d)]. The exact results of the 3D harmonic model suggest a remedy for this unphysical behavior. Namely, one finds (see Appendix \ref{app:exact}) that the renormalization of the mass in the direction perpendicular to the in-plane field,
\be
\frac{1}{m_\perp(\Phi^2 \ll 1)} \approx \frac{1}{m_\perp(0)}\left( 1-\Phi^2\right),
\label{eq:mass1}
\ee
which we derived by arriving at Eq.~\eqref{eq:Hainter}, is in the opposite limit replaced by
\be
\frac{1}{m_\perp(\Phi^2 \gg 1)} \approx  \frac{1}{m_\perp(0)}\left(\frac{1}{1+\Phi^2}\right).
\label{eq:mass2}
\ee
Since Eq.~\eqref{eq:mass1} is the Taylor expansion of Eq.~\eqref{eq:mass2} for $\Phi^2\ll 1$, replacing the former by the latter will improve the overall
accuracy  of the effective 2D model. Explicitly, the replacement in Eq.~\eqref{eq:Hainter} should be
\be
\Phi^2 \to \left(1-\frac{1}{1+\Phi^2} \right).
\label{eq:replacement}
\ee
We find that, interestingly, with this substitution the energies of the 2D model become {\it exactly equal} to the energies of the full 3D model if the magnetic field is purely in-plane and the confinement potential is harmonic. Once one of these conditions is not valid, the energies of the two models are no more identical (see Fig.~\ref{fig:compar_add} for an illustration). Nevertheless, we expect that the two-dimensional effective model with the replacement in Eq.~\eqref{eq:replacement} is a quantitatively reliable representation of the energy effects of the in-plane magnetic field of arbitrary direction and magnitude and for a general heterostructure profile.\footnote{It is conditioned on the assumption that the out-of-plane component of the magnetic field is not very large, meaning it does not destroy the hierarchy of the energies $E_z^* \gg E_x^* \sim E_y^*$, which is the regime of interest for us.}

\section{Discussion}

\label{sec:discussion}

We have derived an effective two-dimensional model which quantitatively describes the orbital effects of the in-plane field on the spectra of quantum dots created in a 2DEG. The corresponding Hamiltonian reads as
\be
H_{\rm 2D}^{\rm eff} = \frac{(\Pin \cdot \Binunit )^2}{2m} +\frac{[\Pin \cdot (\Binunit \times \z ) ]^2}{2m \,\, (1+\Phi^2)} +V_{\rm 2D}(\Rin),
\label{eq:main result}
\ee
where the kinetic momentum $\Pin$ is given in Eq.~\eqref{eq:Pin}, the flux $\Phi$ in Eq.~\eqref{eq:phi}, and the in-plane unit vectors $\Binunit$ and $\Binunit \times \z $ are parallel and perpendicular, respectively, to the in-plane component of the magnetic field $\Bin$. For $\Phi^2 \to 0$, Eq.\eqref{eq:main result} reduces to Eq.~\eqref{eq:H2D}, corresponding to a quasi-two-dimensional electron gas description.

The use of this Hamiltonian is two-fold. If the applied fields are such that the orbital effects can not be neglected and have to be incorporated into the description, it is a substantial simplification if one can still use a 2D model, compared to a fully 3D description. On the other hand, and certainly more importantly, these effects should be taken as a tool to probe quantum dot and its single-particle orbitals. As we have demonstrated, the directional variation of the eigenstate energy gives  direct access to the corresponding orbital shape, that is the size and the orientation with respect to the crystallographic axes. In addition, looking at the same variation as a function of the field magnitude allows one to find the  effective width of the 2DEG, and in turn the microscopic parameters of the interface. For example, for the triangular confinement of a heterostructure, this would be the interface electric field, which in turn allows one to determine the spin-orbit constants.

\begin{figure}
\includegraphics[width=0.49\columnwidth]{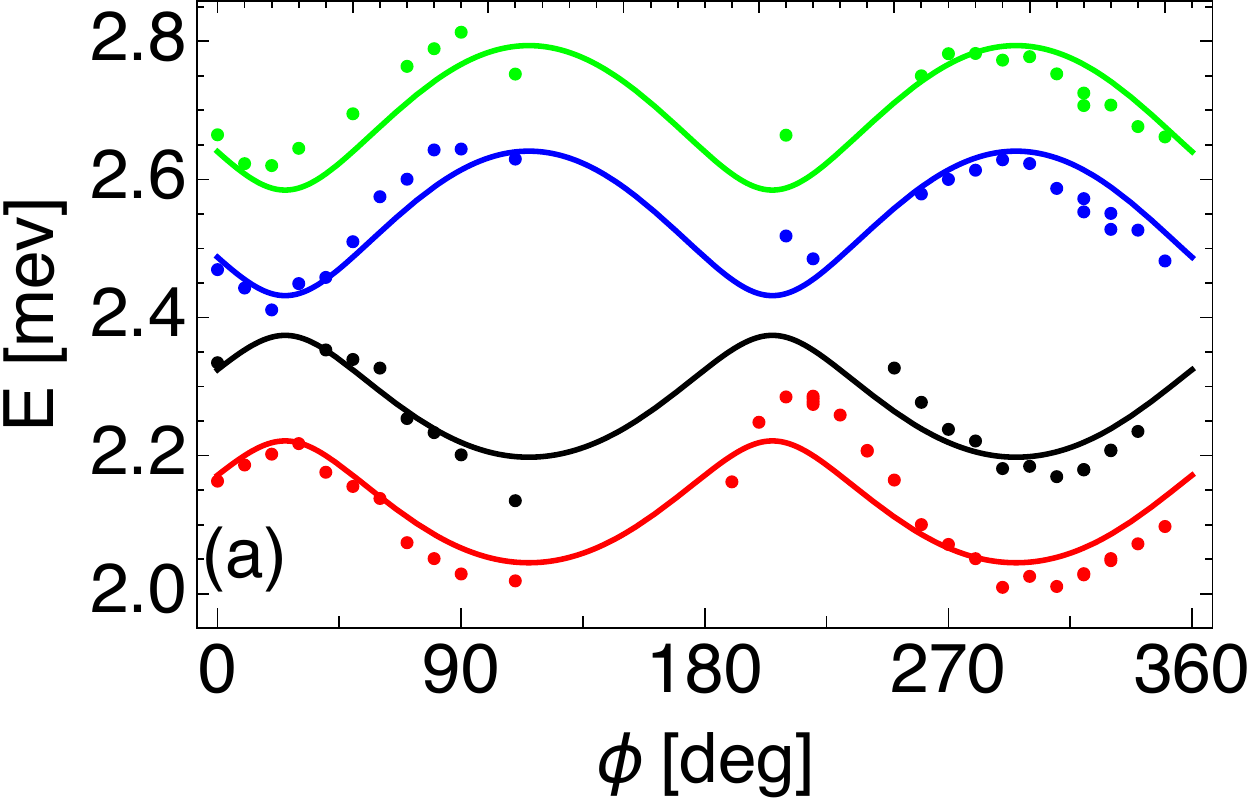}
\includegraphics[width=0.49\columnwidth]{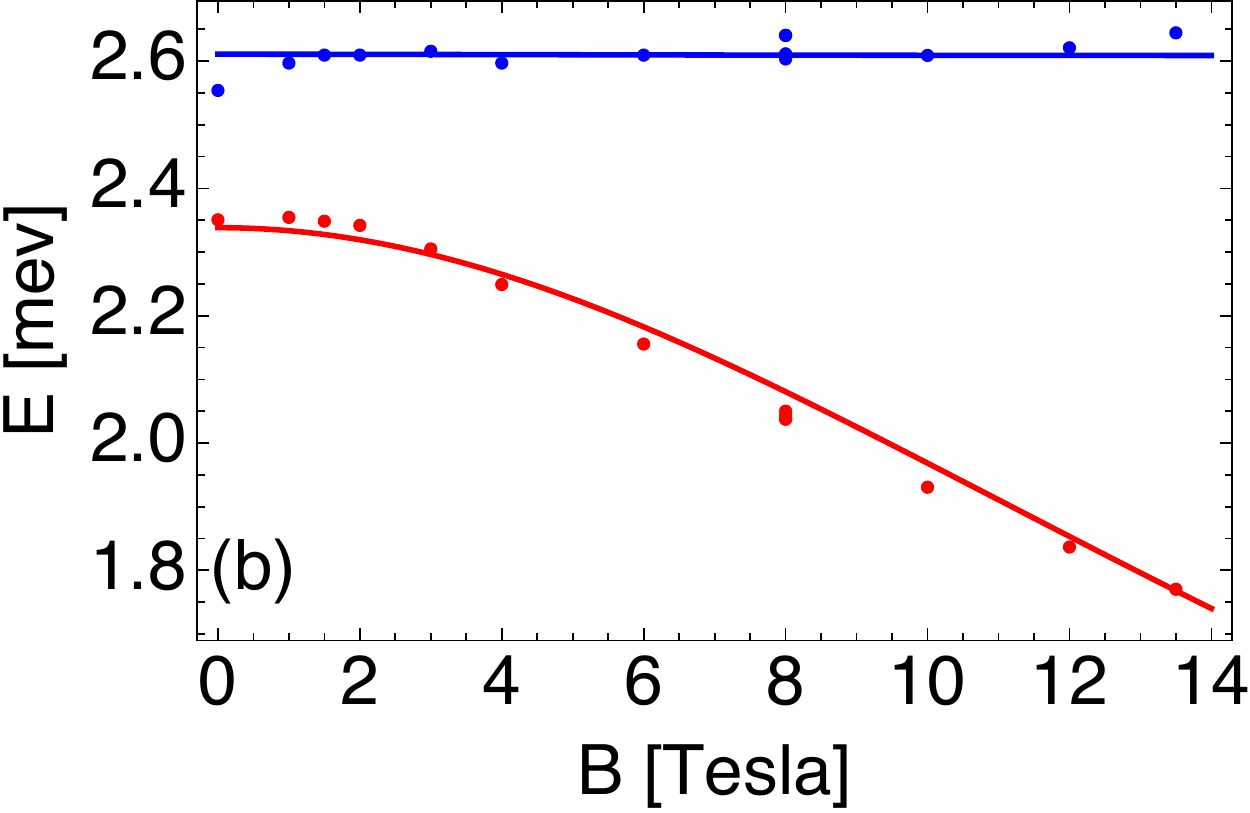}
\caption{\label{fig:data}
The two lowest excitation energies measured in the experiment of Ref.~\citenum{camenzind}, fitted to the exact eigenvalues of $H_{\rm 2D}^{\rm eff}$ [Eq.~\eqref{eq:main result}] (see Appendix \ref{app:exact}). The values of the fitting parameters are given with the error of the last digit in brackets. (a) The directional variation at $B = 8$ T resulted in $\lambda_z=7.19(5)$ nm, $\delta=25(2)^\circ$, $\delta E_x^* = 2.51(1)$ meV, $\delta E_y^* = 2.71(1)$ meV. The red (black) denotes the excitation from the ground state to the lowest orbital without (with) a spin flip. Similarly for the blue and green for the second excited orbital. Here, a constant Zeeman energy is included in the fit for the spin-flip terms, resulting in the $g$ factor $|g|=0.33(2)$.
(b) The field magnitude dependence at $\phi=\delta+\pi/2$ gave $\lambda_z=6.49(5)$ nm, $\delta=51(3)^\circ$, $\delta E_x^* = 2.338(6)$ meV, $\delta E_y^* = 2.611(6)$ meV. Converting the value of $\lambda_z$ to electric field using Fig.~\ref{fig:lambdaz} gives $E_{\textrm{ext}}=2.14(4)$ V/$\mu$m. Here, each point is the average of a Zeeman split pair. We note that the data in panel (a) and (b) were obtained in different cool downs, which might be the reason for the difference in the extracted parameters, especially $\delta$.}
\end{figure}

We illustrate these possibilities on the data measured in the experiment of Refs.~\citenum{yu} and \citenum{camenzind}. We fit the data to the model in Eq.~\eqref{eq:main result}
and plot the result in Fig.~\ref{fig:data}. Figure \ref{fig:data}(a) shows the directional variations of the excitation energies at $B = 8$ T. The data clearly demonstrate that the dot was modestly anisotropic and its main confinement-potential axis was along $\delta \approx 25^\circ$ with respect to the crystallographic [100] axis. Figure \ref{fig:data}(b) shows the excitation energies as a function of the magnetic-field magnitude. Compared to Fig.~\ref{fig:data}(a), this is a more suitable measurement to determine the effective 2DEG width. The fitted value
$\lambda_z \approx 6.5$ nm gives, using Fig.~\ref{fig:lambdaz} (or Table \ref{tab:lambdaz}), appropriate for a heterostructure with a triangular potential, the interface electric field $E_{\textrm{ext}} \approx 2.14$ V/$\mu$m. With this value specified, we now use the standard results of the ${\bf k} \cdot {\bf p}$ theory for the spin-orbit strengths (using the notation of Ref.~\citenum{stano2005:PRB}; see Appendix \ref{app:SOI} for details):
\begin{subequations}
\label{eq:sois}
\begin{eqnarray}
\frac{\hbar^2}{2m l_{br}} &\equiv& \alpha_{br} = \alpha_0 e E_{\textrm{ext}} + (\beta_B-\beta_A) \ave{\delta(z)}{\alpha},\\
\frac{\hbar^2}{2m l_{d}} &\equiv& \alpha_d = \frac{\gamma_c}{\hbar^2} \ave{p_z^2}{\alpha}.
\end{eqnarray}
\end{subequations}
Using $\alpha_0 = -4.7$\AA$^2$, $\beta_B-\beta_A=-1.22$ eV\AA$^2$, and $\gamma_c=-10.6$ eV\AA$^3$ gives the spin-orbit lengths $l_{br}\approx 2.64$ $\mu$m, and $l_{d}\approx3.63$ $\mu$m (ignoring the overall minus sign for both interactions). This translates into the spin-orbit mixing angle $\vartheta=36^\circ$, and the overall scale $l_{so}=2.14$ $\mu$m. Here, $\vartheta$ is defined by $\tan\vartheta = \alpha_{d}/\alpha_{br}$. An {\it independent} fit based on the spin relaxation time anisotropy gave $\vartheta=31^\circ$ and $l_{so}=2.13$ $\mu$m.\cite{camenzind} Alternatively, assuming that the relaxation data give a reliable value for the angle $\vartheta=31^\circ$, while the interface electric field is extracted reliably by the fit shown in Fig.~\ref{fig:data}(b), we can estimate the value for the parameter $\gamma_c$ from these two values and Eqs.~\eqref{eq:sois}. This procedure results in $\gamma_c=-8.8$ eV\AA$^3$, in good agreement with typical values in GaAs obtained by alternative methods.\cite{dettwiler2017:PRX}

\section{Conclusions}

We have analyzed the orbital effects of the magnetic field applied in the plane of a 2DEG, observable in the spectrum of a gated quantum dot. In the leading order, these effects can be succinctly described as an anisotropic renormalization of the electron mass tensor. The renormalization arises due to the finite width of the 2DEG, and depends on the flux corresponding to the magnetic field penetrating the area given as the square of the  effective 2DEG width. We have related this width to common types of heterostructure-interface potentials in detail necessary for a quantitative analysis. Most importantly, the effects allow one to extract the size and orientation of the quantum dot single-particle orbitals, as well as the 2DEG width, thus providing new  characterization methods for gated quantum dots. We illustrated the usefulness of the method by fitting the strengths of the spin-orbit interactions, the linear Rashba, the linear Dresselhaus, and the cubic Dresselhaus terms, from the data measured in Ref.~\citenum{camenzind}.

\acknowledgments
This work was supported by JSPS Kakenhi Grant No. 16K05411, and CREST JST (JPMJCR1675),
the Swiss National Science Foundation (Switzerland), by the NCCR QSIT, the Swiss Nanoscience Institute (SNI) and the European Microkelvin Platform (EMP).

\appendix

\section{Heterostructure potential eigenstates and matrix elements}

We give here, for reference, the energies and some matrix elements of the heterostructure eigenstates which are needed in the main text.

\label{app:matele}

\subsection{Triangular confinement}

\begin{figure}
\includegraphics[width=0.45\columnwidth]{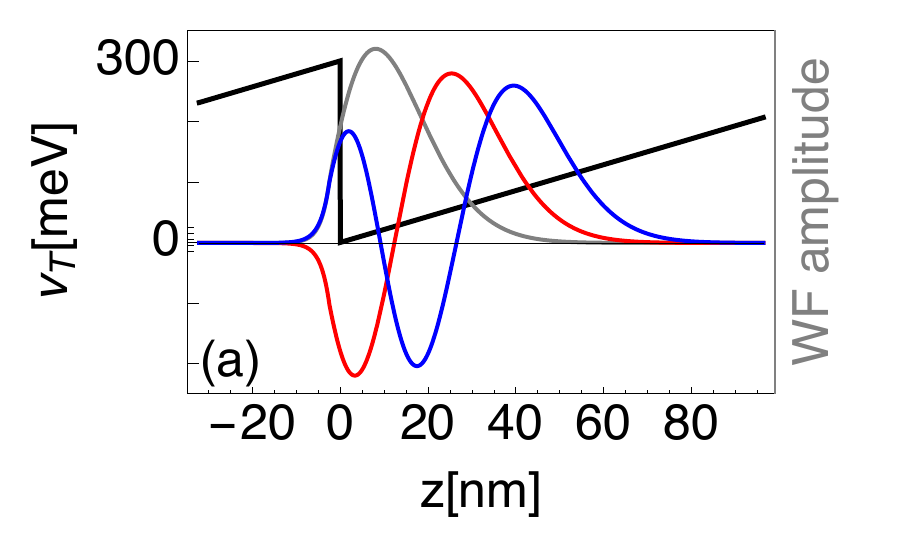}
\includegraphics[width=0.45\columnwidth]{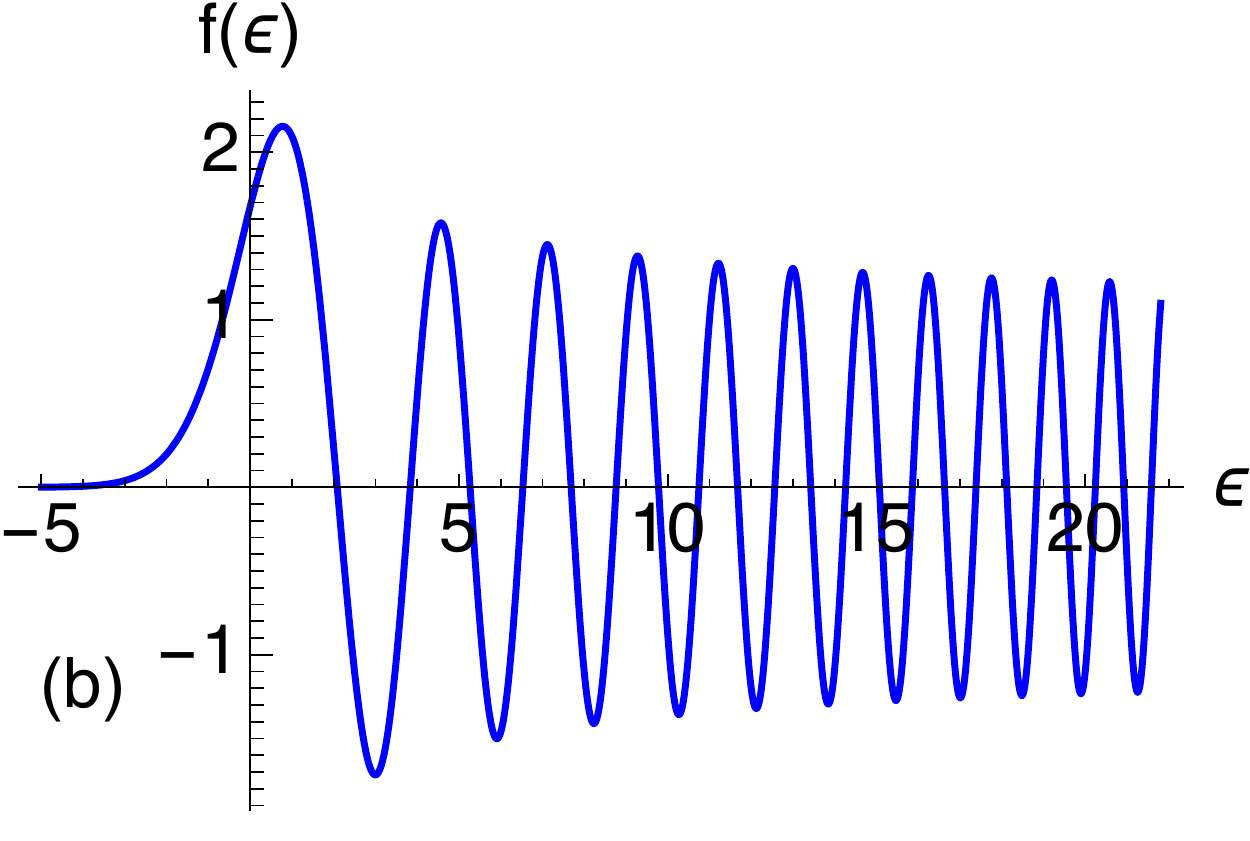}
\includegraphics[width=0.45\columnwidth]{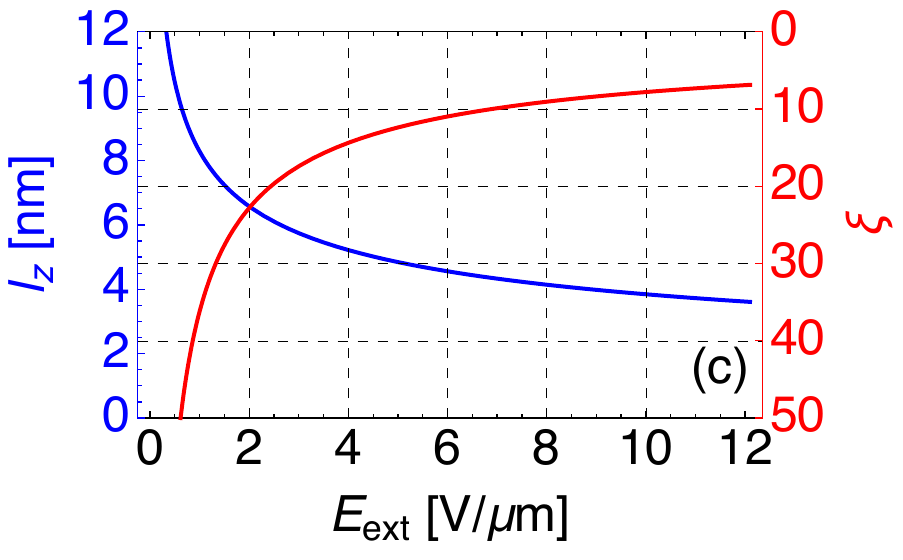}
\includegraphics[width=0.45\columnwidth]{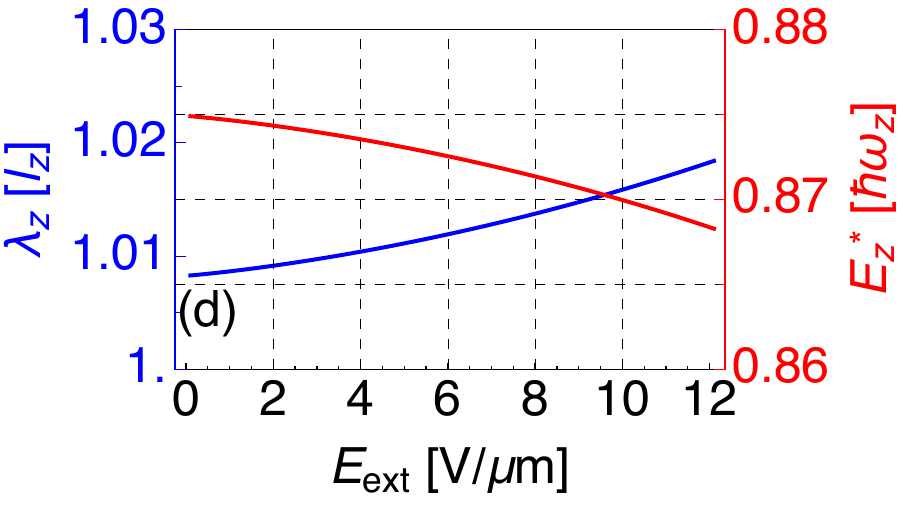}
\includegraphics[width=0.45\columnwidth]{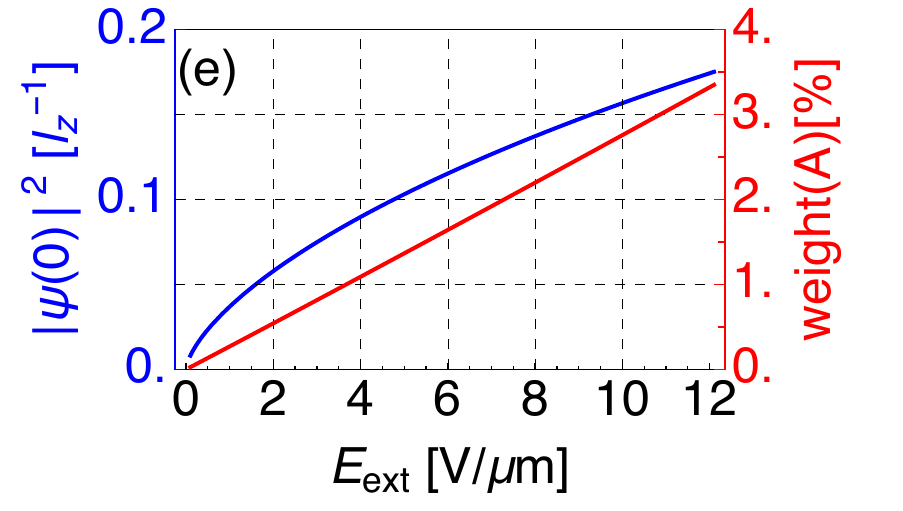}
\includegraphics[width=0.45\columnwidth]{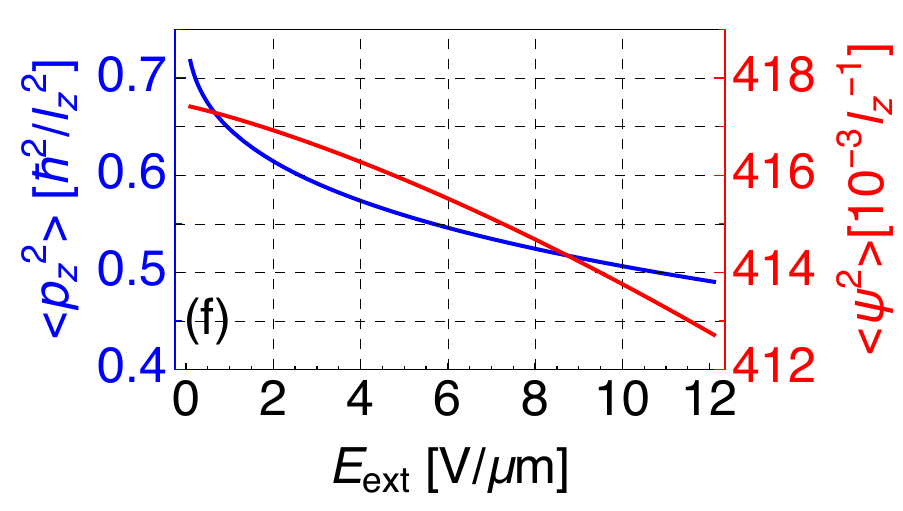}
\caption{\label{fig:triangular}
Illustrations and parameters of the triangular confinement potential model. The band offset is $V_0=300$ meV, the interface electric field is $E_{\textrm{ext}}=2.14$ V/$\mu$m, $A=$ Al$_x$Ga$_{1-x}$As with $x=0.3$, and $B=$ GaAs, unless stated otherwise. (a) The potential profile (black) and the amplitudes of the three lowest wave functions. (b) The function $f(\epsilon)$, roots of which define the allowed energies. (c) The nominal length $l_z$ and the factor $\xi$ as a function of the interface electric field. (d) The effective length and the subband excitation energy in their natural units as a function of the interface electric field. (e) The ground-state wave function density at the interface and its weight in the half-space $z<0$. (f) The expectation value of $p_z^2$ and $\psi(z)^2$ in the ground state.}
\end{figure}

We consider the potential shape as drawn by the black line in Fig.~\ref{fig:triangular}(a). It represents the spatial dependence of the bottom of the conduction band of a heterostructure. It displays a finite offset at $z=0$, due to a different material composition to the left and right of this point, and a linear slope (an electric field) possibly due to remote doping by impurities. In solving for the eigenstates, we neglect the potential variation for $z<0$ and assume that the linear growth for $z>0$ extends to infinity, by which we arrive at Eq.~\eqref{eq:VT}. These simplifications lead to small effects on the quantities of our interest.

With this, the Schr\"odinger equation is
\be
\left(-\frac{\partial}{\partial z}\frac{\hbar^2}{2m(z)}\frac{\partial}{\partial z} +v_T(z) -E \right) \psi(z)=0,
\label{eq:A1}
\ee
where we allow for a position dependence of the effective mass, which takes different values on the two sides of the interface,
\be
m(z) = \left\{
\begin{tabular}{ll}
$m_A$,\,&if $z<0$,\\
$m_B$,\,&if $z>0$.\\
\end{tabular}
\right.
\label{eq:effective mass}
\ee
We solve Eq.~\eqref{eq:A1} in the left and right half of the space separately using the ansatz
\be
\psi(z) = N_A \psi_A(z) + N_B \psi_B(z),
\ee
with the matching conditions
\begin{subequations}
\begin{eqnarray}
N_A \psi_A(0) &=& N_B \psi_B(0),\\
m_A^{-1} N_A \partial_z\psi_A(0) &=& m_B^{-1} N_B \partial_z\psi_B(0).
\end{eqnarray}
\end{subequations}
For $z<0$ the potential is constant, so that
\be
\psi_A(z) = \exp\left[ z \left(\frac{2 m_A (V_0 -E)}{\hbar^2} \right)^{1/2} \right].
\ee
For $z>0$ the equation is
\be
\left(-\frac{\hbar^2}{2m_B}\frac{\partial^2}{\partial z^2} +e E_{\textrm{ext}} z -E \right) \psi(z)=0.
\ee
Introducing a dimensionless length $s=z/l_z$, we get
\be
\left(-\frac{\partial^2}{\partial s^2}+ \frac{2 m_B e E_{\textrm{ext}} l_z^3}{\hbar^2} s -\frac{2m_Bl_z^2 E}{\hbar^2} \right) \psi(s)=0.
\ee
We choose $l_z$ such that the linear term prefactor is 1:\cite{batke1986:PRB}
\be
l_z=\left( \frac{\hbar^2}{2 m_B e E_{\textrm{ext}}} \right)^{1/3},
\label{eq:lzapp}
\ee
and introduce the dimensionless energies $\epsilon=2E/\hbar \omega_z$, and $\xi=2V_0/\hbar \omega_z$, with $\hbar \omega_z=\hbar^2/m_Bl_z^2$. With one more dummy variable, $x=s-\epsilon$, the Schr\"odinger equation takes the form of the Airy differential equation,
\be
\frac{\partial^2}{\partial x^2} y(x) - x y(x) =0.
\ee
The solutions are the Airy functions $\textrm{Ai}(x)$. Using the solutions normalizable at $x\to \infty$, we have
\be
\psi_B(z)=\textrm{Ai}(s-\epsilon).
\ee
Using explicit formulas, the matching conditions read as
\begin{subequations}
\begin{eqnarray}
N_A &=& N_B \textrm{Ai}(-\epsilon),\\
N_A \sqrt{\frac{m_B}{m_A}(\xi - \epsilon)} &=& N_B \textrm{Ai}^\prime(-\epsilon),
\end{eqnarray}
\end{subequations}
and can be written as the quantization condition for the allowed energy values,
\be
f(\epsilon) \equiv \sqrt{\frac{m_B}{m_A}(\xi - \epsilon)}\, \textrm{Ai}(-\epsilon) - \textrm{Ai}^\prime(-\epsilon) =0.
\ee
This function is plotted in Fig.~\ref{fig:triangular}(b), with each root $\epsilon<\xi$ corresponding to a subband. Once the energy is specified, the normalization constant follows as
\be
N_B^{-2} = l_z \left( \sqrt{\frac{m_B}{m_A}} \frac{\textrm{Ai}^2(-\epsilon)}{2\sqrt{\xi-\epsilon}} + \int_{-\epsilon}^\infty  \textrm{Ai}^2(x) {\rm d}x \right).
\ee
For parameters typical for GaAs/AlGaAs heterostructures, for example, $V_0=300$ meV and $E_{\textrm{ext}}$ several Volts per micrometer, the parameter $\xi\gg 1$. In this case, one can find useful results in the limit $\xi\to \infty$  (which also makes the value of $m_A$ irrelevant): $\epsilon_1\approx 1.17 \, \hbar \omega_z$, $\epsilon_2\approx 2.04 \, \hbar \omega_z$, $\lambda_z \approx 1.03 \,l_z$ (the lowest excited subband contributing by 94.3\%), var$_{\alpha=1}(z)\approx 0.486 \, l_z^2$, var$_{\alpha=2}(z)\approx 1.485 \, l_z^2$, $E^*_z \approx 0.875 \, \hbar \omega_z$, and $\ave{p_z^2}{\alpha=1} \approx 0.78 \hbar^2/l_z^2$. Some of these quantities are plotted as functions of the interface electric field on Figs.~\ref{fig:triangular}(c) to \ref{fig:triangular}(f).

\subsection{Rectangular confinement}

\begin{figure}
\includegraphics[width=0.45\columnwidth]{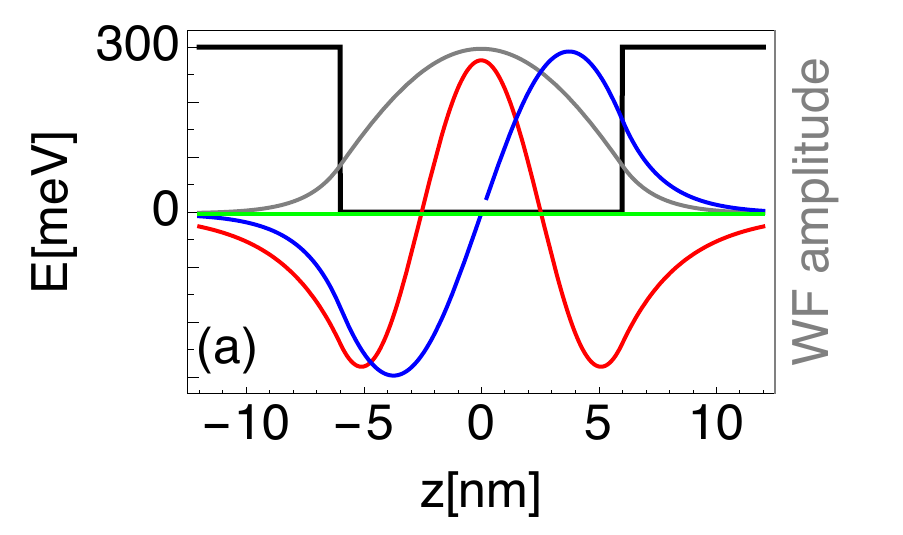}
\includegraphics[width=0.45\columnwidth]{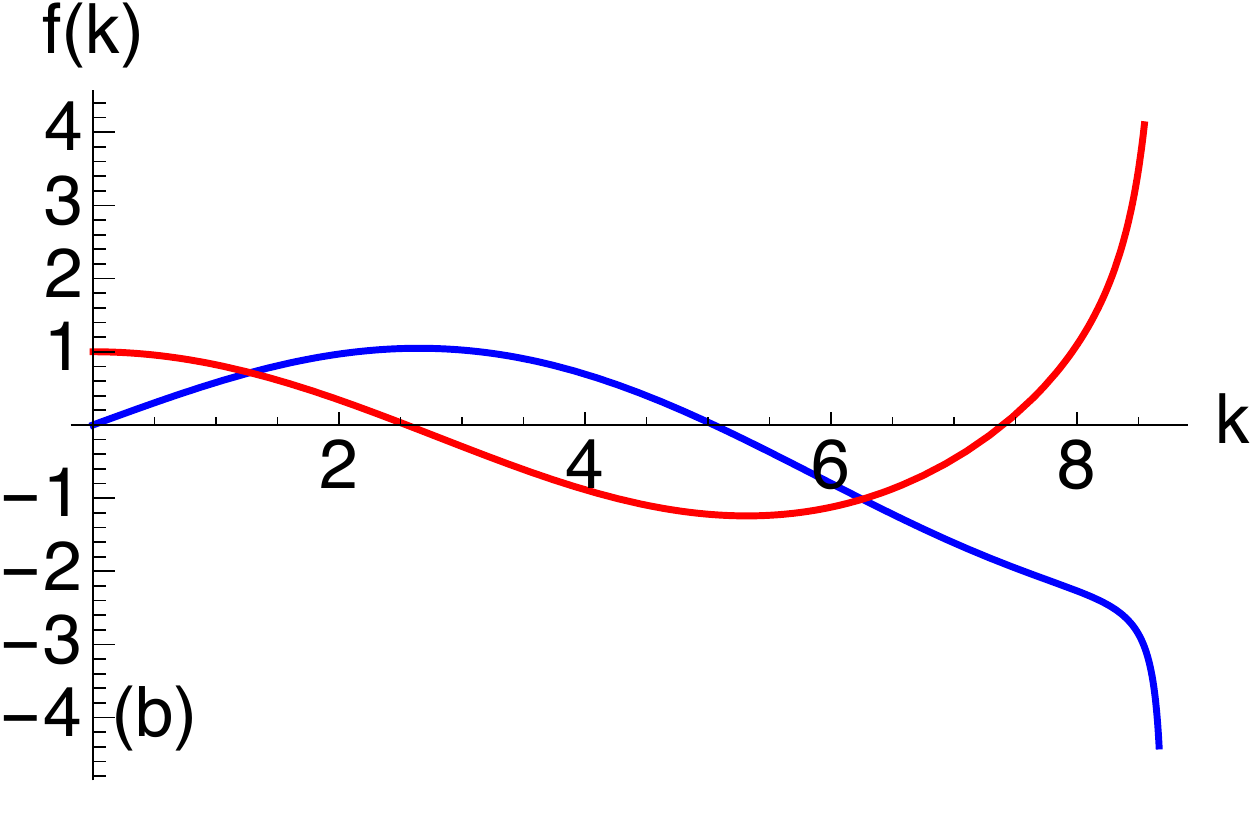}
\caption{\label{fig:rectangular}
Rectangular confinement potential model. The band offset is $V_0=300$ meV and the nominal quantum-well width is $l_z=12$ nm. (a) The potential profile (black) and the amplitudes of the three lowest wave functions. (b) The function $f(k)$, roots of which define the allowed energies for the symmetric solutions (red) and antisymmetric ones (blue).
}
\end{figure}

We obtain the eigenstates in a way analogous to the previous section. Since now the potential is piecewise constant, see Fig.~\ref{fig:rectangular}(a), we skip the details being a textbook quantum mechanics and only give results. The solutions have definite inversion symmetry with respect to $z=0$. Inside the well they take form of the trigonometric functions, $\cos(k z /l_z)$ and $\sin(k z /l_z)$, respectively. Figure \ref{fig:rectangular}(b) shows two functions, the roots of which specify the allowed wavevectors $k$ and the corresponding energies $E(k)=\hbar^2 k^2/2m$. For $m_A=m_B$ and in the limit $V_0 \to \infty$, the solutions become $k=(2n+1)\pi$ for the symmetric subbands and $k=2n \pi$ for the antisymmetric subbands, with $n$ an integer. It leads to $E_z^*/\hbar \omega_z = 3\pi^2/2$, the dipole moment between the lowest two subbands equal to $16 l_z /9 \pi^2$,
$\lambda_z^4 = l_z^4(15-\pi^2)/12\pi^4$, var$_{\alpha=1}(z) = (1/12-1/2\pi^2)\, l_z^2$, var$_{\alpha=2}(z)=(1/12-1/8\pi^2)\, l_z^2$, and $\ave{p_z^2}{\alpha=1} = \pi^2 \hbar^2/l_z^2$. Some of these values are given in Table \ref{tab:lambdaz}.

\subsection{Harmonic confinement}

The matrix elements of the eigenstates of a harmonic potential are obtained from the standard representation of the operators
\begin{subequations}
\begin{eqnarray}
z = \sqrt{\frac{\hbar}{2 m \omega_z}} (a^\dagger +a),\\
p_z = i\sqrt{\frac{\hbar m \omega_z}{2 }} (a^\dagger -a),
\end{eqnarray}
\end{subequations}
with $\hbar \omega_z = \hbar^2/m l_z^2$. The results are given in Table \ref{tab:lambdaz}.

\section{Exact spectrum of a bilinear Hamiltonian}

\label{app:exact}

There are several methods to diagonalize a Hamiltonian which is a quadratic function of coordinates $r_1,r_2, \ldots, r_d$ and  momenta $p_1,p_2, \ldots, p_d$ in any dimension $d$.\cite{rebane1969,davies1985,lin2002} We follow the method used in Refs.~\citenum{rebane1969,schuh1985}, which is based on solving for the unknown operator $L$, linear in $r$'s and $p$'s, which fulfills the equation $[H,L]=\epsilon L$. This can be formulated as an eigenvalue problem, by constructing a $2d$ by $2d$ matrix composed of $2 \times 2$ blocks, where the $(ij)$-th block for $i,j=1,2,\ldots, d$ is defined as
\begin{equation}
\Omega_{ij}=i \hbar \left(
\begin{tabular}{cccc}
$\frac{\partial^2 H}{\partial p_i \partial r_j}$ & $\frac{\partial^2 H}{\partial p_i \partial p_j}$ \\
$-\frac{\partial^2 H}{\partial r_i \partial r_j}$ & $-\frac{\partial^2 H}{\partial r_i \partial p_j}$ \\
\end{tabular}
\right).
\end{equation}
The eigenvalues of matrix $\Omega$ come in pairs, $\{+\epsilon_i,-\epsilon_i\}_{i = 1, \ldots, d}$ and give the $d$ characteristic energies $\epsilon_i$, the excitation energies of the $d$ linear harmonic oscillators.

\subsection{Exact spectrum of the 2D effective model}

Here we are interested in using the above described procedure for the effective 2D model, which treats the in-plane magnetic-field effects perturbatively. This means that $d=2$, and the Hamiltonian is the sum of $H_{\rm 2D}$ [Eq.~\eqref{eq:H2D}] and $H_{\rm inter}^{(\alpha)}$ [Eq.~\eqref{eq:Hainter}]. It results in the following matrix $\Omega$:
\begin{widetext}
\begin{equation}
\Omega=i \hbar \left(
\begin{tabular}{cccc}
0 & $\frac{1-\Phi^2 \cos^2(\delta-\phi)}{m}$ & $-\frac{\omega_c}{2}$ & $-\frac{\Phi^2\sin(2\delta-2\phi)}{2m}  $ \\\\
$- m (\omega_x^2 + \frac{\omega_c^2}{4})$ & 0 & 0 &  $-\frac{\omega_c}{2}$ \\\\
$\frac{\omega_c}{2}$ & $-\frac{\Phi^2\sin(2\delta-2\phi)}{2m}  $ & 0 & $\frac{1-\Phi^2 \sin^2(\delta-\phi)}{m}$   \\\\
0 & $\frac{\omega_c}{2}$ & $- m (\omega_y^2 + \frac{\omega_c^2}{4})$ & 0
\end{tabular}
\right),
\end{equation}
\end{widetext}
where we denoted $\hbar \omega_c = \hbar e B_z/m$. The characteristic equation for the eigenvalues $\epsilon$ of $\Omega$ is
\be
\epsilon^4 - b \epsilon^2 +c=0,
\ee
where
\begin{subequations}
\begin{eqnarray}
b &=& \hbar^2 \omega_x^2+\hbar^2 \omega_y^2+\hbar^2\omega_c^2-\Phi^2(A^2+\hbar^2 \omega_c^2/4),\\
c &=& (1-\Phi^2) \hbar^4 \omega_x^2 \omega_y^2 - \Phi^2 A^2 \hbar^2 \omega_c^2/4,
\end{eqnarray}
\end{subequations}
and we introduced a confinement anisotropy related parameter
\be
A^2 = \hbar^2 \omega_x^2 \cos^2(\delta-\phi) + \hbar^2 \omega_y^2 \sin^2(\delta-\phi).
\ee
The two solutions for the energies are given by
\be
\epsilon^2_{1,2} = \frac{b \pm \sqrt{b^2-4c}}{2}.
\ee
By Taylor expanding the previous equation in parameter $\Phi^2$, and setting $B_z=0$, we obtain
\begin{subequations}
\begin{eqnarray}
\epsilon_1 &=& \hbar \omega_x [1-\Phi^2 \cos^2(\delta-\phi)] + O(\Phi^{4}),\\
\epsilon_2 &=& \hbar \omega_y [1-\Phi^2 \sin^2(\delta-\phi)]+ O(\Phi^{4}),
\end{eqnarray}
\end{subequations}
which gives Eqs.~\eqref{eq:dEstar}. Similarly, doing a Taylor expansion in $(\hbar\omega_x-\hbar \omega_y)$, gives Eqs.~\eqref{eq:dEstar2}.

\subsection{Exact spectrum of the 3D harmonic model}


We now consider the 3D model with a harmonic confinement in all three directions, that is the one described by Eqs.~\eqref{eq:V2D} and \eqref{eq:VH}. The energies can be obtained by a straightforward analogy of the previous subsection applied for $d=3$. We do not repeat the explicit formulas, as they were given in Ref.~\citenum{rebane2012} as Eqs.~(6), (14), (17), and (18) therein. Using these, we derive the in-plane energies for a symmetric in-plane potential $\omega_x = \omega_y$ and a purely in-plane field. In the limit $\Phi^2 \ll 1$ we get
\begin{subequations}
\begin{eqnarray}
\epsilon_1 &=& \hbar \omega_x ,\\
\epsilon_2 &=& \hbar \omega_x \sqrt{1-\Phi^2},
\end{eqnarray}
\end{subequations}
while in the opposite limit $\Phi^2 \gg 1$ we have
\begin{subequations}
\begin{eqnarray}
\epsilon_1 &=& \hbar \omega_x ,\\
\epsilon_2 &=& \hbar \omega_x \frac{1}{\sqrt{1+\Phi^2}}.
\end{eqnarray}
\end{subequations}
This gives Eqs.~\eqref{eq:mass1} and \eqref{eq:mass2}.

\begin{table}
\begin{tabular}{cccccc}
\hline \hline
parameter & $E_0$  & $\Delta_0$ & $P_0$ & $m$ & $E_c$\\
\hline
unit & eV & eV& eV\AA& $m_e$ & eV\\
\hline
GaAs & 1.519 & 0.341 & 9.88 & 0.067 & 0\\
AlGaAs & 3.13 & 0.3 & 8.88 & 0.150& 1.12\\
\hline \hline
\end{tabular}
\caption{\label{tab:parameters}
The band-structure parameters used in Appendix C (see Ref.~\citenum{acta}, p.~688). The band gap $E_0$, the split-off energy $\Delta_0$, the interband matrix element $P_0$, the effective mass $m$, the conduction band offset $E_c$.
We interpolate the parameters for Al$_{1-x}$Ga$_x$As by linear interpolation using the doping $x=0.3$, except for the band-structure offset, where we use the approximation $E_c$(Al$_{1-x}$Ga$_x$As) $ \approx 0.773\, x$ meV (see Appendix 3, p.~412, in Ref.~\citenum{davies}).
}
\end{table}

\section{Spin-orbit strengths}

\label{app:SOI}

To estimate the strengths of the Rashba spin-orbit interactions, we use formulas from Ref.~\citenum{acta} (see p.~679--681 therein). The heterointerface electric field $E_{\textrm{ext}}$ contributes by
\be
\alpha_{br}^{(1)} = \alpha_0 e E_{\textrm{ext}},
\ee
with (Ref.~\citenum{acta} Eq.~III.105)
\be
\alpha_0=\frac{P_0^2}{3}\left( \frac{1}{(E_0 +\Delta_0)^{2}} - \frac{1}{E_0^{2}}\right).
\label{eq:alpha0}
\ee
For the parameters of GaAs, see Table \ref{tab:parameters}, and the electric field $E_{\textrm{ext}}=2.14$ V$/\mu$m, it gives
\be
\alpha_{br}^{(1)} \approx -1.0\,\textrm{meV\AA}.
\label{eq:abr1}
\ee
Using a slightly different prefactor, $\alpha_0=-5.15$ \AA$^2$, from Ref.~\citenum{knap1996:PRB}, we would get
\be
\alpha_{br}^{(1)} \approx - 1.1\, \textrm{meV\AA}.
\ee
The abrupt change in the band-structure parameters at the heterostructure interface contributes by (Ref.~\citenum{acta}, Eq.~III.106)
\be
\alpha_{br}^{(2)} = (\beta_B - \beta_A) \langle {\delta(z)} \rangle,
\ee
where (Ref.~\citenum{acta}, Eq.~III.98)
\be
\beta = \frac{P_0^2}{3}\left( \frac{1}{E_0 +\Delta_0-E_c} - \frac{1}{E_0-E_c}\right).
\label{eq:beta}
\ee
Using the model described in Appendix \ref{app:matele}, for $E_{\rm ext}=2.14$ V/$\mu$m we get the wave-function density at the interface $|\psi_{\alpha=1}(0)|^2\approx 0.06/l_z$, out of which approximately 25\% is contributed by the difference in the effective mass (not shown). With this
\be
\alpha_{br}^{(2)} \approx -1.15\, \textrm{meV\AA}.
\label{eq:abr2}
\ee
The Dresselhaus term is given by
\be
\alpha_{d} = \frac{\gamma_c}{\hbar^2} \langle {p_z^2} \rangle.
\ee
Using again Appendix \ref{app:matele} we have $\langle p_z^2\rangle \approx 0.61/l_z^2$ which, together with $\gamma_c=-10.6$ eV\AA$^3$, finally gives
\be
\alpha_{d} \approx - 1.57\, \textrm{meV\AA}.
\label{eq:ad1}
\ee
This value, together with $\alpha_{br}=-2.15$ meV obtained from Eqs.~\eqref{eq:abr1} and \eqref{eq:abr2}, was used in Eq.~\eqref{eq:sois} in Sec.~\ref{sec:discussion}.

\end{document}